\documentclass[12pt]{iopart2}
\usepackage{iopams}
\usepackage{epsfig}
\newcommand{\be}{\begin{equation}}
\newcommand{\ee}{\end{equation}}
\newcommand{\bd}{\begin{displaymath}}
\newcommand{\ed}{\end{displaymath}}
\newcommand{\BE}{\begin{eqnarray}}
\newcommand{\EE}{\end{eqnarray}}

\newcommand{\bra}{\left\langle}
\newcommand{\ket}{\right\rangle}
\newcommand{\order}{{\cal O}}

\newcommand{\id}{{\rm 1\!I}}

\newcommand{\bx}{\ensuremath{\mathbf{x}}}

\newcommand{\avg}[1]{\left\langle{#1}\right\rangle}

\newcommand{\bC}{\ensuremath{\mathbf{C}}}
\newcommand{\bD}{\ensuremath{\mathbf{D}}}

\newcommand{\bG}{\ensuremath{\mathbf{G}}}
\newcommand{\bGz}{\ensuremath{\mathbf{G_0}}}

\newcommand{\bR}{\ensuremath{\mathbf{R}}}

\newcommand{\bta}{\mbox{\boldmath $\eta$}}
\newcommand{\btap}{\mbox{\boldmath $\eta'$}}
\newcommand{\btapp}{\mbox{\boldmath $\eta''$}}

\newcommand{\hbx}{\widehat{\mbox{\boldmath $x$}}}

\newcommand{\boldpsi}{{\mbox{\boldmath $\psi$}}}

\newcommand{\tp}{t^\prime}

\begin{document}

\title[Dynamics of random replicators with Hebbian interactions]{Dynamics of random replicators with Hebbian interactions}

\author{Tobias Galla\dag\ddag
}

\address{\dag\ The Abdus Salam International Centre for Theoretical Physics, Strada Costiera 11, 34014 Trieste, Italy}
\address{\ddag\ INFM/CNR, Trieste-SISSA Unit, V. Beirut 2-4, 34014 Trieste, Italy}

\begin{abstract}
A system of replicators with Hebbian random couplings is studied using
dynamical methods. The self-reproducing species are here characterized
by a set of binary traits and interact based on complementarity. In
the case of an extensive number of traits we use path-integral
techniques to demonstrate how the coupled dynamics of the system can
be formulated in terms of an effective single-species process in the
thermodynamic limit, and how persistent order parameters
characterizing the stationary states may be computed from this process
in agreement with existing replica studies of the statics. Numerical
simulations confirm these results. The analysis of the dynamics allows
an interpretation of two different types of phase transitions of the
model in terms of memory onset at finite and diverging integrated
response, respectively. We extend the analysis to the case of
diluted couplings of an arbitrary symmetry, where replica theory is
not applicable. Finally the dynamics and in particular the approach to
the stationary state of the model with a finite number of traits is
addressed.
\end{abstract}

\pacs{87.23-n, 87.10+e, 75.10.Nr, 64.60.Ht}

\ead{\tt galla@ictp.trieste.it}

\section{Introduction}
The dynamics and collective phenomena of interacting agents in models
of game theory, economy and biology are currently studied intensively
in the physics community for example in the context of the so-called
Minority Game \cite{Book1,Book2,Book3}. In particular the tools
of equilibrium and non-equilibrium statistical mechanics have proved
to be extremely powerful for the analysis of such complex systems with
many degrees of freedom. Another prominent example of such systems
consists in so-called replicator models (RM) which are commonly used
to describe the evolution of self-reproducing interacting species
competing within a given framework of limited resources. RM have found
wide applications in a variety of fields including game theory,
socio-biology, pre-biotic evolution and optimization theory
\cite{Book4,Book5}. While earlier studies were mostly restricted to the
deterministic case the first model with random interactions, the
so-called random replicator model (RRM), was introduced by Diederich
and Opper in \cite{DO,OD}. The randomness of the interspecies
couplings here reflects an amount of uncertainty about the structure
of interactions in real ecosystems. The initial RRM and various
extensions have subsequently been studied in a series of papers
\cite{OD2,OF1,OF2,OF3,O,SF,Tokita, ParisiBiscari,TokitaYasumoti,ChaTok} and they have been found to exhibit intriguing features, both from the biological
point of view as well as from the perspective of statistical
mechanics. In particular it has been realised that the model exhibits
phase behaviour with interesting ergodic and non-ergodic phases,
different types of phase transitions as well as replica-symmetry
breaking.

From the point of view of statistical mechanics, RRMs can be seen as
disordered complex systems with quenched random couplings. The
variations of RRMs mentioned above then differ in the details of the
statistics from which the couplings are drawn. In the original model
of Diederich and Opper the interactions between the species were
assumed to be pairwise and drawn from a Gaussian distribution,
\cite{OF1} discusses the case of higher-order Gaussian interactions
and \cite{OF3}, \cite{O} and
\cite{SF} are concerned with models in which the couplings are of a
separable Hebbian structure, reminiscent of those used extensively in
the context of neural networks \cite{Cool00a,Cool00b}.

Most of this existing work on RRM in the physics literature seems to
focus on either numerical simulations and/or on the application of
static methods borrowed from statistical mechanics, most notably
replica techniques. The only exception in which exact dynamical
techniques are employed to deal with the disordered interactions of
RRM appears to be the early paper \cite{OD} by Opper and Diederich
(and the follow-up
\cite{OD2}) and the work by Rieger \cite{Rieger}, where RRM are addressed using dynamical generating
functionals \`a la De Dominicis \cite{dD}. The use of replica methods
to study the statics of such models relies on the existence of a
global Lyapunov (or fitness) function, the stationary states of the
replicator system are then identified as extrema of this random
function. This, however, constitutes a severe constraint from the
point of view of biology as it limits the accessible systems to those
with symmetric couplings. Models with (even partial) asymmetry in the
interactions do not allow one to formulate the dynamics in terms of a
Lyapunov function so that replica techniques can not be applied in
this case. Hence, under such circumstances a direct study of the
dynamics of the systems is required.
 
In this paper we discuss a dynamical approach to a class of RRM with
Hebbian interactions. In the fully connected case the inter-species
interactions are purely symmetric so that we can compare the results
we obtain from a generating functional approach to earlier analyses of
the statics of the same model
\cite{O,SF}. In particular we show that both methods lead to the same
equations for the observables describing the stationary states. The
study of the dynamics however enables us to shed some additional light
on the two different types of phase transitions previously observed in
the statical analysis of the model, and allows for an interpretation
of these transitions in terms of an instability of the fixed point
reached, and equivalently in terms of ergodicity breaking, dependence
on initial conditions and of memory onset.  We then extend the
analysis to cases of Hebbian couplings with dilution of an arbitrary
symmetry, where replica theory is no longer applicable. Finally a
discussion of the dynamics of the model with a finite number of traits
is presented. Here we can go beyond the computation of the statistics
of fixed-point solutions and can make statements regarding the approach
to the stationary state.

\section{Model definitions}
The basic ingredients of the model are as follows. We consider a
population of $N$ species, labelled by Roman indices $i=1,\dots,N$,
and with concentrations $x_i\in[0,\infty)$. The system evolves in time
according to the following replicator equations
\be\label{eq:dyn}
\frac{d}{dt} x_i(t)=-x_i(t)\left[\sum_{j=1}^N \frac{c_{ij}}{c}J_{ij} x_j(t)+\sigma\zeta_i(t)-\lambda(t)\right].
\ee
Here $J_{ij}$ denotes the coupling between species $i$ and $j$,
$J_{ij}<0$ corresponds to a pair of co-operating species, whereas $i$
and $j$ are competing with each other, if $J_{ij}>0$. The
$c_{ij}\in\{0,1\}$ specify the dilution of the interactions, they are drawn such that any given link $c_{ij}$ is present with probability $c\in(0,1]$, further details of their distribution will be given below. Following
\cite{O,SF} each species $i$ is then assumed to carry $P$ traits,
$\xi_i^\mu$, labelled by an index $\mu=1,\dots,P$. In the first part
of the paper we will take the number of traits to be proportional to
the system size, i.e. $P=\alpha c N$, where $\alpha={\cal O}(1)$ is an
additional model parameter. The case $P={\cal O}(1)$ is discussed in
the final section before the conclusions. The traits $\{\xi_i^\mu\}$
take values $\{-1,1\}$ with equal probability and are drawn
independently before the dynamics is started and then remain fixed.
The interaction between species is then given by the Hebb rule
\be
J_{ij}=\frac{1}{N}\sum_{\mu=1}^{\alpha c N} \xi_i^\mu\xi_j^\mu,\quad i\neq j,
\ee
 Two species $i$ and $j$ are the more co-operative, the more
 complementary traits $\mu$ with $\xi_i^\mu\xi_j^\mu=-1$ they have. If
 $i$ and $j$ carry a large number of identical traits, they will be
 competitors.  The traits $\{\xi_i^\mu\}$ along with the dilution
 variables $\{c_{ij}\}$ represent the quenched disorder of the
 problem. The latter are drawn from a distribution with the following
 statistics ($i\neq j$):
\be
c=\avg{c_{ij}}_c=\avg{c_{ji}}_c, \quad \avg{c_{ij}c_{ji}}_c-c^2=\Gamma c(1-c) 
\ee
($\avg{\dots}_c$ stands for an average over the dilution
variables). $c\in(0,1]$ denotes the connectivity, we here focus on the
case of extensively many connections per species ($cN\sim{\cal
O}(N)$). The parameter $\Gamma$ controls the symmetry of the dilution,
for $\Gamma=1$ one has $c_{ij}=c_{ji}$, whereas $c_{ij}$ and $c_{ji}$
are uncorrelated for $\Gamma=0$. Choosing $0\leq\Gamma\leq 1$ allows
one to interpolate between these two cases. For $c=1$ the parameter
$\Gamma$ becomes irrelevant and we recover the fully connected model
studied by static means in \cite{O,SF}. We set $c_{ii}=1$ and the
diagonal elements $J_{ii}$ are fixed to
\be
J_{ii}=2u,
\ee
where $u\geq 0$ stands for the self-interaction, or so-called
co-operation pressure \cite{OD,Book4}. The scaling of the
self-interaction with $c$ is chosen similar to the choice of
\cite{SF}\footnote{The analysis can be adapted to the case
$(c_{ii}J_{ii})/c=2u={\cal O}(c^0)$ upon replacing $u\to uc$
throughout the calculation.}  and accordingly we do not address the
so-called `extremely dilute limit' $\lim_{N\to\infty} c= 0$ while
still $\lim_{N\to\infty} cN=\infty$ sometimes considered in the
context of neural networks \cite{Cool00a,Cool00b}. $u$ is an
additional model parameter, large values of the self-interaction will
favour states in which all species concentrations remain positive
whereas for small $u$ the disordered non-diagonal terms $J_{ij}$
dominate the interactions and some fraction of the species may die out
asymptotically. The $\{\zeta_i(t)\}$ in (\ref{eq:dyn}) represent
Gaussian white noise of zero mean and unit variance, so that the
concentration of each species is subject to random fluctuations of
magnitude $\sigma$. We will mostly concentrate on the case $\sigma=0$
but consider $\sigma>0$ to study the stability of the solutions of the
noiseless dynamics. Finally, the Lagrange parameter
$\blambda=\{\lambda(t)\}$ is chosen to ensure the normalization
\be\label{eq:norm0}
\sum_{i=1}^N x_i(t)=N
\ee
at all times, so that the total concentration of species is kept constant as the eco-system evolves in time. We note that in absence of the noise $\zeta_i(t)$ Eq. (\ref{eq:dyn}) may be written as
\be
\frac{d}{dt} x_i(t)=x_i(t)\left[f_i(t)+\lambda(t)\right],
\ee
where $f_i(t)=-\sum_{j} (c_{ij}/c)J_{ij}x_j(t)$ is the `fitness' of species $i$ at time $t$. The normalization constant $\lambda(t)$ is then related to the `mean fitness' by
\be
\lambda(t)=-\frac{1}{N}\sum_j x_j(t) f_j(t).
\ee
In terms of game theory the variables $i$ can be understood to label
different strategies and $x_i(t)$ as the fraction of individuals
playing strategy $i$ at time $t$ in a given population of agents . A
player playing strategy $i$ in an ensemble characterized by the
vector $\bx=(x_1,\dots,x_N)$ then receives a payoff $f_i(\bx(t))$. Any
individual $i$ is the assumed to reproduce at a rate proportional to
the difference between $f_i(t)$ and the mean fitness across the
population at time $t$. All offsprings of individual $i$ then inherit
the strategy of $i$.
\section{Generating functional analysis}

\subsection{Generating functional and single effective species process}
In the thermodynamic limit the time-evolution of the $N$ coupled
replicator equations (\ref{eq:dyn}) is conveniently studied using
generating functional techniques originally developed for disordered
systems by De Dominicis \cite{dD}. These allow one to express dynamics in
terms of macroscopic order parameters such as the correlation and
response functions of the system and have proved to be extremely
powerful not only for the analysis of spin-glasses and neural
networks, but for example also in the context of the Minority Game and
other agent-based models \cite{Book2}; generating functionals have
also been used to study simples cases of RRM in
\cite{OD,Rieger}. In order to generate response functions we first add
perturbation fields $\{h_i(t)\}$ to the replicator dynamics and start from
\be\label{eq:dyn2}
\frac{d}{dt} x_i(t)=-x_i(t)\left[\sum_{j=1}^N \frac{c_{ij}}{c}J_{ij} x_j(t)+\sigma\zeta_i(t)-\lambda(t)+h_i(t)\right].
\ee 
The generating functional is then defined as
\BE
Z[\boldpsi]&=&\bra e^{i\sum_i\int dt
x_i(t)\psi_i(t)}\ket_{paths}\nonumber \\ &\equiv&\int D\bx D\hbx\,
p(\bx(0))\exp\left[i\sum_i\int dt \hat x_i(t)\left(\sum_{j=1}^N \frac{c_{ij}}{c} J_{ij}
x_j(t)\right)\right]\nonumber\\ &&\hspace{-1cm}\times
\exp\left[i\sum_i\int dt \hat x_i(t)\left(\dot
y_i(t)-\lambda(t)+\sigma\zeta_i(t)+h_i(t)\right)\right].
\EE
Following \cite{OD} we have here employed a transformation
$y_i(t)=\log x_i(t)$ (for $x_i(t)>0$) in order to bring the replicator
equations into a suitable form. Note that any $x_i$ which is
initialized at $x_i(t=0)=0$ will remain zero forever. The functional
$Z[\boldpsi]$ (with $\boldpsi=\{\boldpsi(t)\}_{t\geq
0}=\{(\psi_1(t),\dots,\psi_N(t))\}_{t\geq 0}$) thus represents the
Fourier transform of the measure describing the probability of a
certain path $\{\bx(t)=(x_1(t),\dots,x_N(t))\}_{t\geq 0}$ to
occur. $p(\bx(0))$ describes the distribution of initial conditions
from which the dynamics is started at time $t=0$. Once the
disorder-averaged generating functional $\overline{Z[\boldpsi]}$ has
been computed, correlation and response functions may be obtained by
taking derivatives $\overline{Z[\boldpsi]}$ of with respect to the
source and perturbation fields $\{\psi_i(t),h_i(t)\}$.
 
The procedure of performing the disorder-average in the thermodynamic
limit $N\to\infty$ is standard \cite{OD,Cool00b, Book2} and results in a
self-consistent closed system of equations for the dynamical order
parameters of the problem. For the present system these order
parameters are given by the Lagrange parameter $\blambda=\{\overline{\avg{\lambda(t)}}\}$ and the
correlation and response functions $\{\bC, \bG\}$:
\be
C(t,t')=\lim_{N\to\infty}\frac{1}{N}\sum_{i=1}^N \overline{\avg{x_i(t)x_i(t')}}, ~~~ G(t,t')=\lim_{N\to\infty}\frac{1}{N}\sum_{i=1}^N \frac{\delta \overline{\avg{x_i(t)}}}{\delta h_i(t')}. 
\ee
Here $\avg{\dots}$ denotes an average over the possibly random initial
conditions and over realisations of the white noise variables
$\{\zeta_i(t)\}$. Note that $G(t,t')=0$ for $t\leq t'$ due to
causality, and that in general we expect the $G(t,t')$ to take
negative values, given the effective sign of the perturbation $h_i(t)$
in (\ref{eq:dyn2}). Unlike in the case of spherical models of
spin-glasses \cite{Crisanti} no explicit closed differential equations
for $\{\blambda, \bC, \bG\}$ can be found in the present
model. Instead the order parameters are to be computed
self-consistently from a non-Markovian stochastic process for an
effective single-species concentration $x(t)$. This process turns out
to be given by
\BE
\frac{d}{dt} x(t)&=&-x(t)\bigg(\frac{2u}{c}x(t)+\alpha\int_{0}^t d\tp [c\bG(\id-\bG)^{-1}+\Gamma(1-c)\bG](t,\tp)x(\tp)\nonumber\\
&&~~~~~~~~~~~~~-\lambda(t)+h(t)+\sigma\zeta(t)+\eta(t)\bigg)\label{eq:effprocess}.
\EE
Here we have assumed that the distribution of initial conditions $p(\bx(0))$ factorizes over the individual species $\{i\}$ and that the perturbation fields $\{h_i(t)\}$ take the form $h_i(t)\equiv h(t)$ for all $i$. $\{\eta(t)\}$ is Gaussian coloured noise with temporal correlations
\be\label{eq:noise}
\alpha\Lambda(t,t')\equiv\avg{\eta(t)\eta(t')}=\alpha[c(\id-\bG)^{-1}\bC(\id-\bG^T)^{-1}+(1-c)\bC](t,t').
\ee
$\id$ denotes the identity matrix, and $\{\zeta(t)\}$ is white noise of zero mean and unit variance and reflects the white noise present in the original multi-species problem. The above correlation and response functions $\bC$ and $\bG$ can in turn be shown to be given by
\be\label{eq:selfcons}
C(t,t')=\avg{x(t)x(t')}_\star,~~~ G(t,t')=\frac{\delta}{\delta h(t')}\avg{x(t)}_\star,
\ee
and $\blambda=\{\lambda(t)\}$ has to be chosen such that 
\be\label{eq:norm}
\avg{x(t)}_\star = 1 \qquad \forall t,
\ee
corresponding to the constraint (\ref{eq:norm0}) in the original
dynamics \footnote{Technically speaking we here only demand that the
relation $\lim_{N\to\infty}N^{-1}\sum_i\overline{\avg{x_i}}=1$ be
satisfied. For $N\to\infty$ the differences between the condition
(\ref{eq:norm0}) imposed on any sample of the original process and
this softer constraint are expected to be irrelevant; see also
\cite{Book2} for a similar case in the context of spherical Minority
Games.}. Here $\avg{\dots}_\star$ denotes an average over realisations
of the effective process (\ref{eq:effprocess}), i.e. over realisations
of the single-species coloured noise $\{\eta(t)\}$ and the white noise
$\{\zeta(t)\}$. The perturbation field $h(t)$ served us only to
generate response functions, and will be set to zero from now on. Note
also that (up to a sign) $G(t,t')$ may equivalently be obtained by
taking a derivative of $\avg{x(t)}_\star$ with respect to
$\lambda(t)$.

Eqs. (\ref{eq:effprocess},\ref{eq:noise},\ref{eq:selfcons},\ref{eq:norm})
 thus form a closed set of equations from which $\{\blambda,\bC,\bG\}$
 are to be obtained. These equations are exact and fully equivalent to
 the original $N$-species problem Eq. (\ref{eq:dyn}) in the limit
 $N\to\infty$. The retarded self-interaction and the coloured noise
 are a direct consequence of the quenched disorder in the original
 problem and impede an explicit analytical solution of the
 self-consistent saddle-point equations for the full two-time objects
 $\bC,\bG$ and the function $\blambda$. In similar disordered systems
 (such as the Minority Game) one therefore has to resort to
 specific ans\"atze for the trajectories of the effective particles
\cite{Book2}. Alternatively, effective dynamical problems in discrete
time may be addressed using a Monte-Carlo integration of the resulting
non-Markovian single-particle processes \cite{EO}. While this method
allows one to compute the dynamical order parameters numerically
without finite-size effects the iteration quickly becomes costly in
terms of computer time as the number of time-steps is increased. For
similar disordered systems one is usually limited to $\order(100)$
iterations, the equilibration times for suitably discretised versions
of the present system however turn out to be much larger.
\subsection{Stationary state}
In order to proceed we assume that a stationary time-translation invariant state is reached in the long-term limit, i.e. we will be looking for solutions of the following type
\be
\lim_{t\to\infty} C(t+\tau,t)=C(\tau), ~~~ \lim_{t\to\infty} G(t+\tau,t)=G(\tau), ~~~ \lim_{t\to\infty}\lambda(t)=\lambda.
\ee
Furthermore we will only address ergodic stationary states, that is states in which perturbations have no long-term effects so that the integrated response function 
\be
\chi\equiv\int_0^\infty d\tau \, G(\tau)
\ee
remains finite, $|\chi|<\infty$, and so that no long-term memory is
present, i.e. we will assume
\be
\lim_{t\to\infty} G(t,t')=0\qquad\forall t'.
\ee
Note again that with the present definitions we expect $\chi$ to be
negative in general, as the perturbation $h_i(t)$ effectively acts
with a negative sign on $\dot x_i$ in Eq. (\ref{eq:dyn2}). At this
stage a specific ansatz needs to be made to address the stationary
states reached by the effective process. We will henceforth restrict
the analysis to the case $\sigma=0$, in which the original problem
(\ref{eq:dyn}) becomes fully deterministic once the disordered
interactions and random initial conditions have been drawn. In this
case it conveniently turns out in simulations that all trajectories
$\{x_i(t)\}$ evolve into fixed points $x_i\equiv
\lim_{t\to\infty} x_i(t)$ for all $i$ for large enough values of $u$\footnote{In the case of symmetric couplings ($c=1$ or $\Gamma=1$) this statement holds for all $u$. This is due to the existence of a Lyapunov function governing the dynamics, and similar behaviour has been observed in RRM with Gaussian symmetric couplings \cite{OD2}. In the case of asymmetric couplings fixed point solutions are generally only found in the regime of large self-interaction $u$, larger than a critical value $u^\star$ which depends on the details of the model \cite{OD2}.}. This observation simplifies the analysis
considerably, as we may now accordingly look for solutions $x(t)\to x$
of the effective process, resulting in a flat correlation function
\be
q\equiv C(\tau)\qquad \forall \tau
\ee
in the stationary state. $1/q$ here serves as a measure of the
diversity of the system, large values of $1/q\approx 1$ indicate a
large fraction of surviving species with roughly equal concentrations
($x_i\approx 1~~\forall i$), for small $1/q\ll 1$ only a small number
of species survives asymptotically in the present model (in the
present model the number of surviving species is extensive); see also
\cite{OF2} for further details of the interpretation of $1/q$.  Within
the fixed point ansatz we also assume that the single-species noise
$\{\eta(t)\}$ approaches a time-independent value $\eta$, which in
turn is a static Gaussian random variable of zero mean and variance
(cf. Eq. \ref{eq:noise}))
\be
\avg{\eta^2}=\alpha \left(c\frac{q}{(1-\chi)^2}+(1-c)q\right).
\ee
We abbreviate $\Sigma^2\equiv\alpha \left(c\frac{q}{(1-\chi)^2}+(1-c)q\right)$ from now on. Any particular realization of $\eta(t)$ in the effective process thus leads to a realization of the random variable $\eta$, which in turn determines the value of the fixed point $x$ reached by that particular trajectory of the effective process. In the following we write $\eta=\Sigma z$ with
$z$ a standard Gaussian variable. Note that for the moment we will only assume the existence of such fixed points, their local stability against perturbations will be discussed in detail below.

Inserting this ansatz into the effective process (\ref{eq:effprocess})
one finds (for $h(t)=0$)
\be\label{eq:fp}
x\left(\frac{2u}{c}x+\alpha\left(c\frac{\chi}{1-\chi}+\Gamma(1-c)\chi\right)x-\lambda+\Sigma z\right)=0,
\ee
and we conclude that the random variable $x$ can take values of either $x=0$ or $x=\frac{\lambda-\Sigma z}{\frac{2u}{c}+\alpha\left(c\frac{\chi}{1-\chi}+\Gamma(1-c)\chi\right)}$, depending on $z$.
Given that $x(t)\geq 0$ the latter solution can only be realised provided that $z<\Delta$, where $\Delta=\lambda/\Sigma$ (the denominator $\frac{2u}{c}+\alpha\left(c\frac{\chi}{1-\chi}+\Gamma(1-c)\chi\right)$ comes out positive), so that we write
\be\label{eq:xofz}
 x(z)=\frac{\lambda-\Sigma z} {\frac{2u}{c}+\alpha c\frac{\chi}{1-\chi}+\Gamma(1-c)\chi} \Theta\left[\lambda-\Sigma z \right],
\ee
following \cite{OD}, with $\Theta(y)$ the step-function,
i.e. $\Theta(y)=1$ for $y>0$ and $\Theta(y)=0$ otherwise. We note here
that $x(z)=0$ is a solution of (\ref{eq:fp}) for all values of $z$. As
it will turn out later (see Eq. (\ref{eq:zstable})) these zero fixed
points are unstable against perturbations if $\lambda-\Sigma z>0$, which justifies (\ref{eq:xofz}) {\em a
posteriori}. Note also that we choose initial conditions such that
$x_i(0)>0$ with probability one for all $i$, as concentrations set to
zero from the start remain zero throughout the dynamics. 

Using (\ref{eq:xofz}) the self-consistency requirements
(\ref{eq:selfcons}) and (\ref{eq:norm}) translate into the following
relations:
\be
q=\avg{x^2}_\star, ~~~ \chi=-\avg{\frac{\partial x}{\partial \lambda}}_\star, ~~~ \avg{x}_\star=1.
\ee
In explicit form these may be written as
\BE
\fl~~~\left[\alpha\left(\frac{cq}{(1-\chi)^2}+(1-c)q\right)\right]^{-1/2}\left[\frac{2u}{c}+\alpha\left(c\frac{\chi}{1-\chi}+\Gamma(1-c)\chi\right)\right] &=&f_1(\Delta), \label{eq:sp1}\\
\fl~~~\left[\alpha\left(\frac{c}{(1-\chi)^2}+(1-c)\right)\right]^{-1}\left[\frac{2u}{c}+\alpha\left(c\frac{\chi}{1-\chi}+\Gamma(1-c)\chi\right)\right]^2&=&f_2(\Delta), \label{eq:sp2} \\
\fl~~~-\left[\frac{2u}{c}+\alpha\left(c\frac{\chi}{1-\chi}+\Gamma(1-c)\chi\right)\right]\chi&=&f_0(\Delta) \label{eq:sp3}.
\EE
Here $f_n(\Delta)=\int_{-\infty}^\Delta Dz (\Delta-z)^n$ ($n=0,1,2$)
with $Dz=\frac{dz}{\sqrt{2\pi}}e^{-z^2/2}$ a standard Gaussian
measure. We note that $\phi=\int_{-\infty}^\Delta Dz=f_0(\Delta)$ represents the
probability of a given species $i$ to survive in the long-term limit,
i.e. to attain a fixed point value $x_i=\lim_{t\to\infty}x_i(t)>0$. We
will refer to $\phi$ as the fraction of surviving species in the
following. For $c=1$ Eqs. (\ref{eq:sp1}, \ref{eq:sp2},
\ref{eq:sp3}) are identical to those found in replica-symmetric studies of the statics of the fully connected model \cite{O,SF}. For $\Gamma=1$ they also coincide with the replica results reported in \cite{SF} for the model with symmetric dilution (up to a suitable re-scaling of $\alpha$ in terms of $c$). Eqs.  (\ref{eq:sp1}, \ref{eq:sp2},
\ref{eq:sp3}) are readily solved numerically  (in terms of $\lambda,\,q$ and $\chi$) for all values of the model parameters.

\begin{figure}[t]
\vspace*{1mm}
\begin{tabular}{cc}
\epsfxsize=72mm  \epsffile{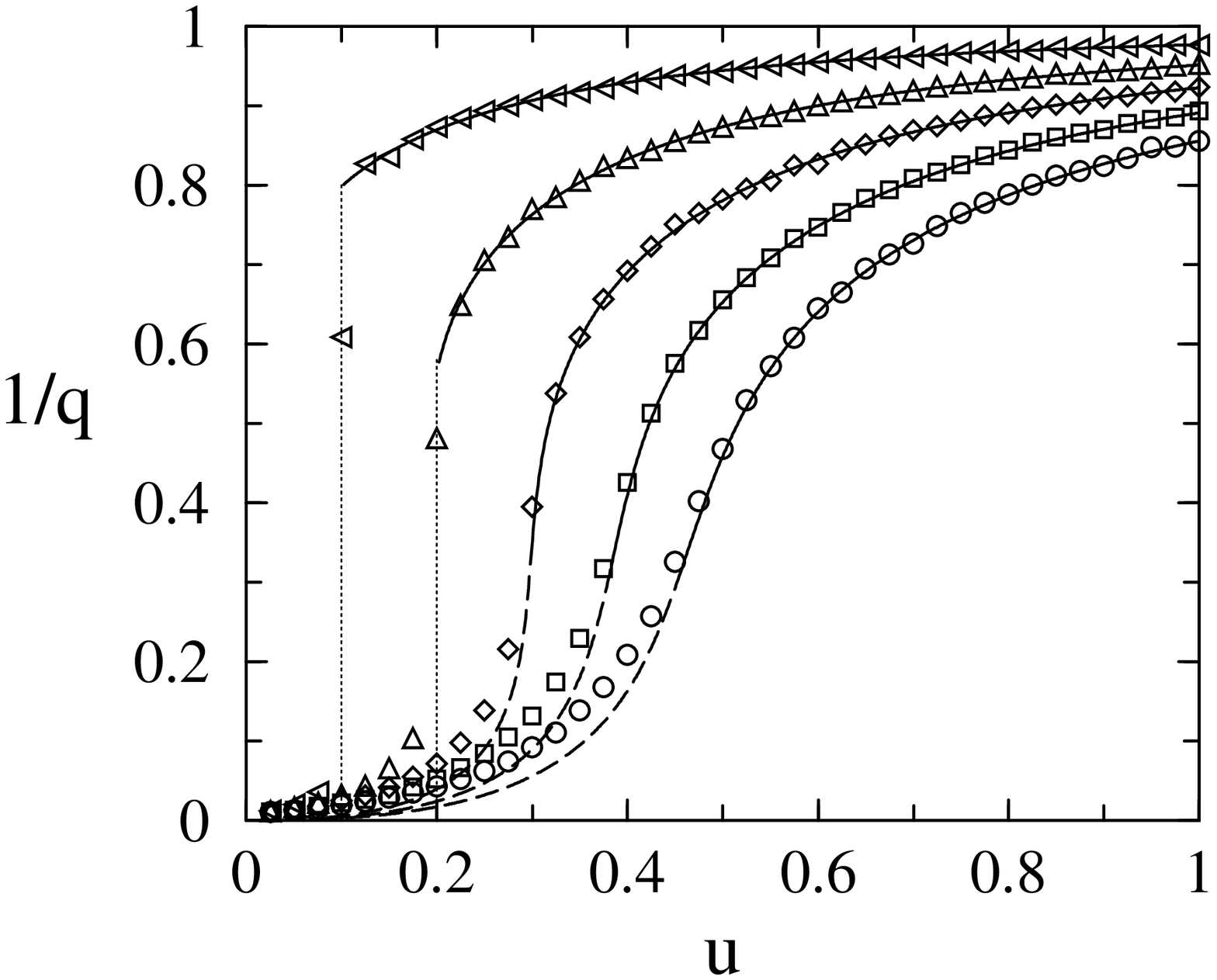} ~&~~
\epsfxsize=72mm  \epsffile{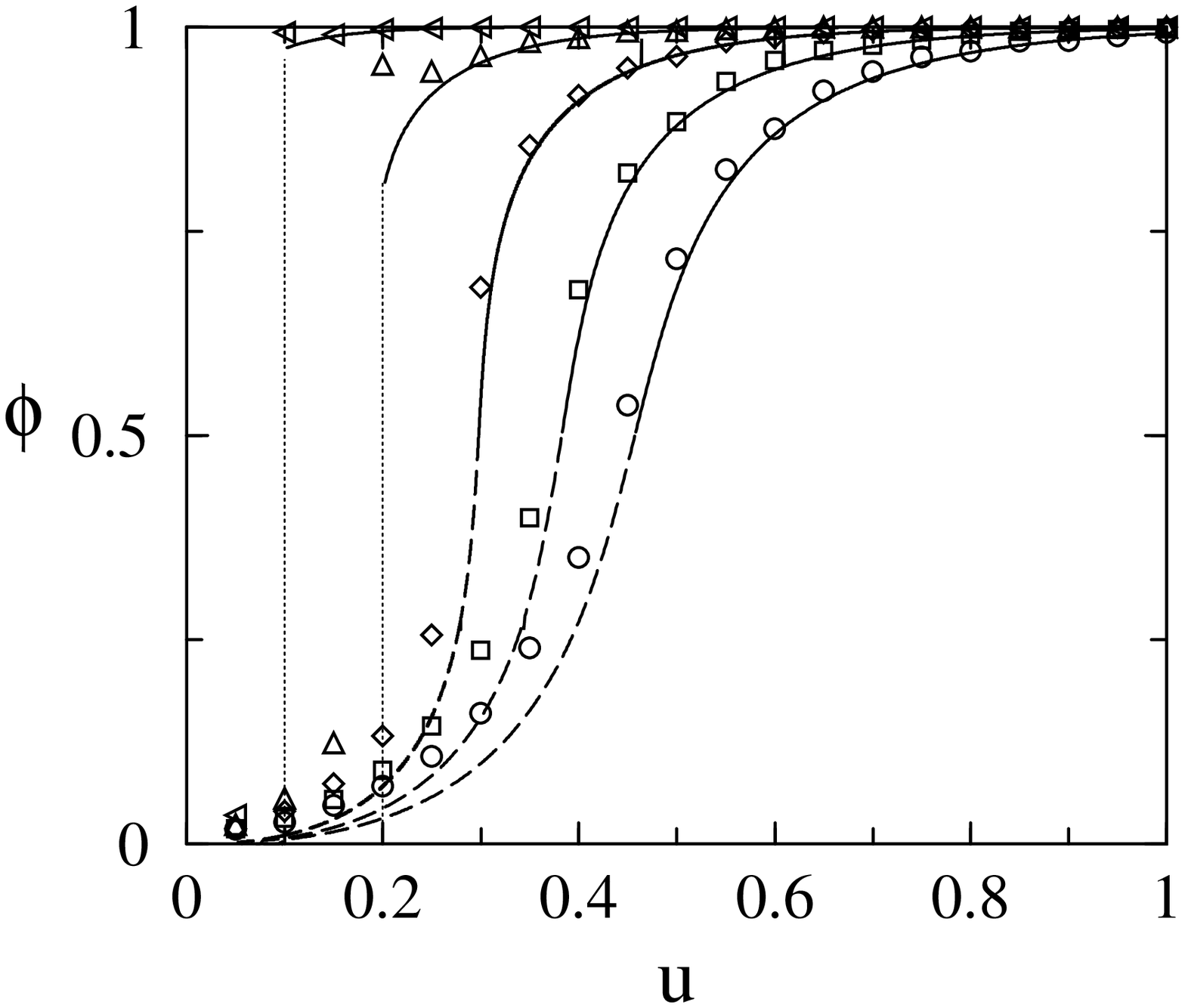}
\end{tabular}
\vspace*{4mm} \caption{Reciprocal of the persistent correlation $q$ and fraction of surviving species $\phi$ vs $u$ for different values of $\alpha$. Solid lines are the analytical predictions for the ergodic phase $u>u^\star(\alpha,c,\Gamma)$. For $\alpha>1/2$ they have been continued into the phase with long-term memory and finite $\chi$ as dashed lines (where the ergodic theory is no longer valid). Vertical dotted lines mark the discontinuous transitions at $u=\alpha/2$ for $\alpha<1/2$. Symbols are from simulations (circles: $\alpha=1$, squares: $\alpha=0.8$, diamonds: $\alpha=0.6$, triangle up: $\alpha=0.4$, left triangle: $\alpha=0.2$). Simulations are performed for $N=300$ species, data are averages over $50$ samples of the disorder ($20$ samples for the fraction of surviving species). The discretisation method used here is that of \cite{OD2}, with effective time-step $\Delta t=\left[c+N^{-1}\sum_i x_i(t)f_i(t)\right]^{-1}$ at time $t$. We here use $c=10$, the discretised dynamics is iterated for $20000$ steps. Dynamics is started from independent random initial conditions drawn from a flat distribution over the interval $x_i(0)\in[0,2]$.}
\label{fig:qphi}
\end{figure}
The resulting predictions for the order parameters $q$ and $\phi$ for
the fully connected case $c=1$ are tested against numerical
simulations of the process (\ref{eq:dyn}) in Fig. \ref{fig:qphi} (the
dilute case will be discussed below) \footnote{Note that extinct
species die out only asymptotically, $\lim_{t\to\infty} x_i(t)=0$, so
that $x_i(t)$ remains positive for all $i$ at finite simulation times
$t$. The fraction of surviving species plotted in Fig. \ref{fig:qphi}
is measured by applying a condition $x_i>\vartheta$ at the end of the
simulation runs. We have tested several values of $\vartheta$ and
slight variations of $\phi$ appear hard to avoid as $\vartheta$ is
varied. For the data in Fig. \ref{fig:qphi} a value of
$\vartheta=0.01$ was used.}, and we observe near perfect agreement in
the regime of large values of $u$, greater than a critical value
$u=u^\star$. The numerical value of $u^\star$ will in general depend
on the model parameters $\alpha,c$ and
$\Gamma$. $u^\star=u^\star(\alpha,c,\Gamma)$ will be determined
analytically in the following section. Below
$u^\star(\alpha,c,\Gamma)$ systematic deviations from the ergodic
theory are observed, the reason for these will become clear below. We
attribute small discrepancies between theory and simulations at values
$u\geq u^\star$ close to the transition points $u^\star$ to
finite-size or finite-running time effects or to numerical errors due
to the discretisation of the original continuous-time replicator
dynamics. We would like to point out that
Eqs. (\ref{eq:sp1})-(\ref{eq:sp3}) admit another (unphysical) solution
in the fully connected case for $\alpha<1/2$ and intermediate values
of $u$, which are not shown in the figure.

\section{Memory onset and phase diagram}

\begin{figure}[t]
\vspace*{1mm}
\begin{tabular}{cc}
\epsfxsize=72mm  \epsffile{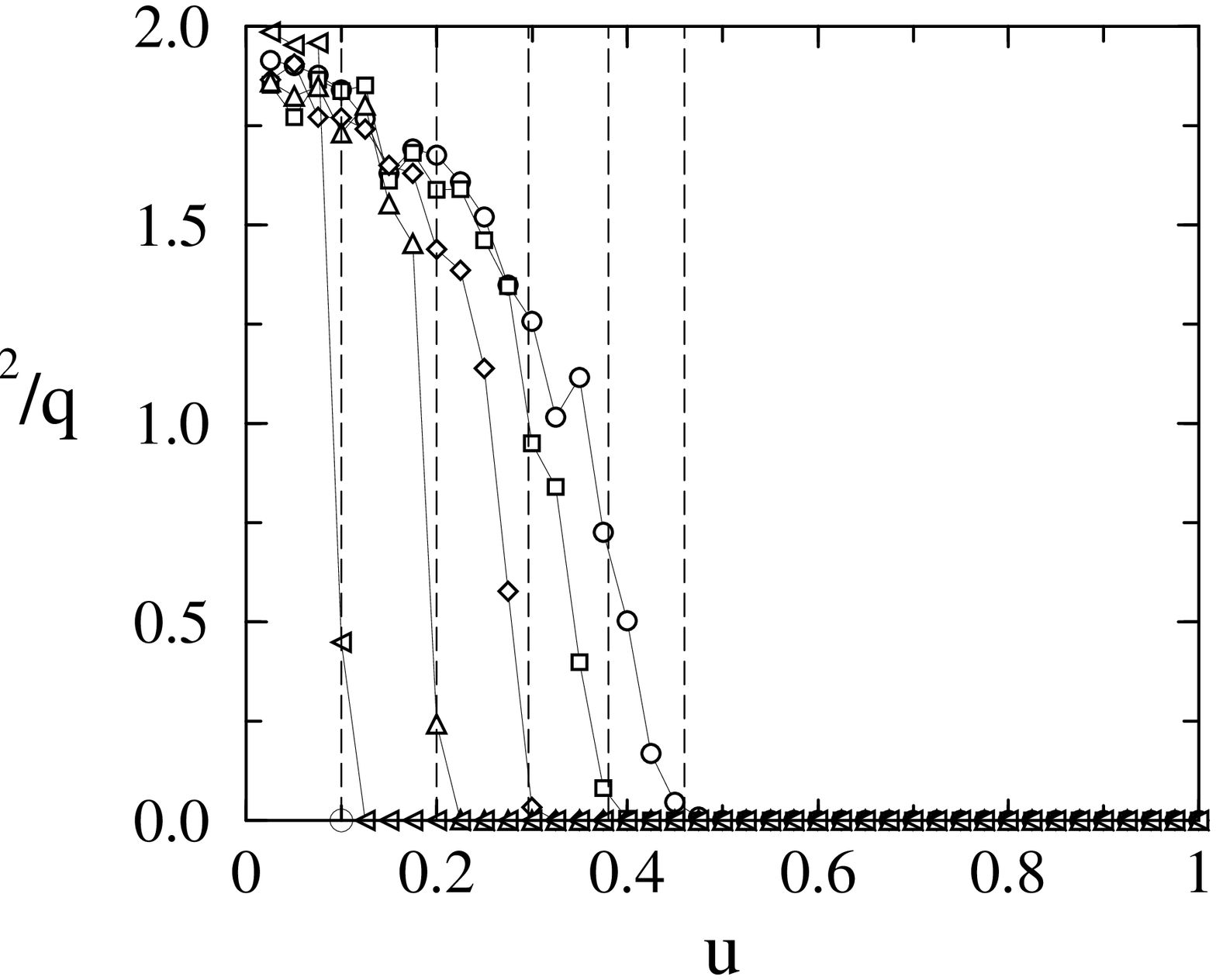} ~&~~
\epsfxsize=72mm  \epsffile{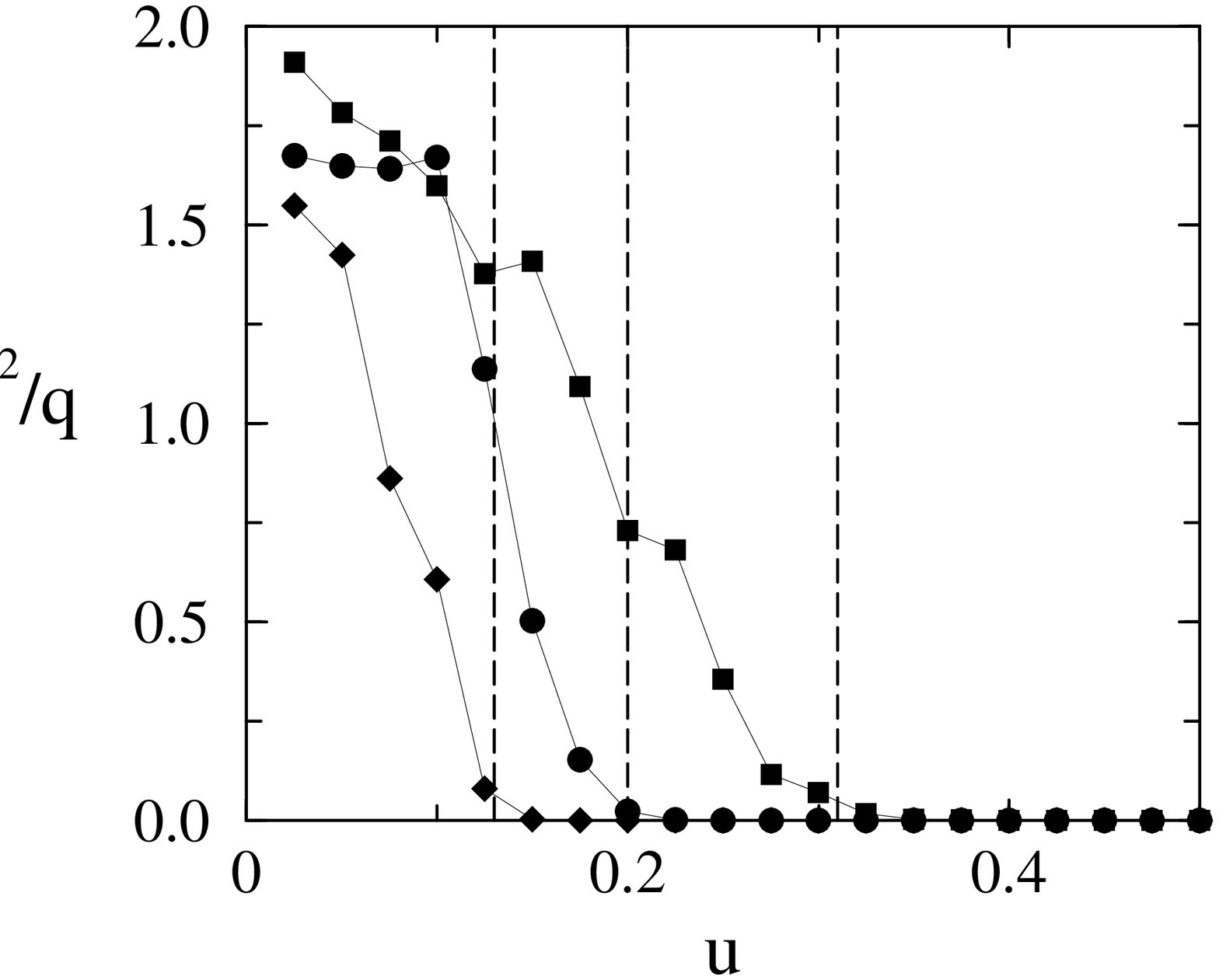}
\end{tabular}
\caption{Reduced distance $d^2/q$ between stationary states obtained for a fixed realization of the disorder, but started from independent initial conditions. The connected markers represent results from simulations. Left panel: fully connected model (circles: $\alpha=1$, squares: $\alpha=0.8$, diamonds: $\alpha=0.6$, triangle up: $\alpha=0.4$, left triangle: $\alpha=0.2$), right panel: model with dilution (with $(\alpha,c,\Gamma)=(1,0.5,0),(1,0.5,1),(1,0.25,0.5)$ for filled circles, squares and diamonds respectively. Simulation parameters as in Fig. \ref{fig:qphi}, $N=400$ species for the dilute systems. The vertical dashed lines in both panels indicate the locations of the ergodic/non-ergodic phase transitions as predicted by the analytical theory.}
\label{fig:distance}
\end{figure}

The ansatz we made to address the stationary states implies an
explicit assumption regarding the ergodicity of the system. In
particular our theory applies only in a regime in which the system
does not exhibit any long term memory and in which stationary state
does not depend on initial conditions. Apart from being locally stable
we therefore require the fixed point reached in the stationary state
to be unique. To test this assumption we have performed
simulations in which one realization of the disorder is fixed and the
system is then started from two different random initial conditions
$\{x_i(0)\}$ and $\{x_i'(0)\}$ (with all $x_i(0), x_i'(0)$ drawn from
a flat distribution over the interval $[0,2]$). We depict the
resulting distance $d^2=N^{-1}\sum_i (x_i-x_i')^2$ between the two
fixed points $\{x_i\}$ and $\{x_i'\}$ reached in the long-term limit
in Fig. \ref{fig:distance} (we normalize by $q$ for convenience). In
an ergodic state without memory we expect this distance to vanish, so
that initial conditions are irrelevant for the stationary state. As
shown in Fig. \ref{fig:distance} this is indeed the case for large $u$
greater than $u^\star(\alpha,c,\Gamma)$, but initial conditions
become relevant below $u^\star$.  This suggests a phase transition
between an ergodic phase at large $u$ and a non-ergodic regime at low
values of $u$. We will now proceed to compute the critical value
$u^\star(\alpha,c,\Gamma)$ separating these two regimes analytically.
\subsection{Diverging integrated response}
A breakdown of our ergodic theory may be signalled by a violation of
either of the assumptions we made to address the stationary states. We
will first inspect the requirement that the integrated response $\chi$
be finite. Taking the limit $\chi\to-\infty$ (recall $\chi<0$) in
Eq. (\ref{eq:sp3}) shows that such a divergence can occur at finite
$u$ only for $\Gamma=0$ or in the fully connected case $c=1$ and directly
leads to $u=\alpha c^2/2$. On the other hand the combination of
Eqs. (\ref{eq:sp2}) and (\ref{eq:sp3}) in the limit $\chi\to-\infty$
implies
\be
\frac{1}{\alpha c}=\frac{f_2(\Delta)}{(f_0(\Delta))^2}.
\ee
Noting that the ratio on the right-hand-side takes only values greater or equal than $2$ (the minimum is attained for $\Delta=0$), we conclude that a divergence of $\chi$ may occur only for $\alpha c\leq 1/2$. 

For $c=1$ such a singularity is indeed observed in the numerical
solutions of Eqs. (\ref{eq:sp1},\ref{eq:sp2},\ref{eq:sp3}), precisely
at the point where discontinuities in the order parameters occur at
$u^\star=\alpha/2$ in our simulations for $\alpha<1/2$, see
Fig. \ref{fig:qphi}. For $\alpha>1/2$ no divergence of the integrated
response is found. In the model with dilution ($c<1$) no such
singularity is observed, for $\Gamma>0$ it is excluded by the
arguments given above, for $\Gamma=0$ it is preceded by a different
type of transition, as we will discuss below.

\begin{figure}
\vspace*{1mm}
\begin{center}
\epsfxsize=80mm  \epsffile{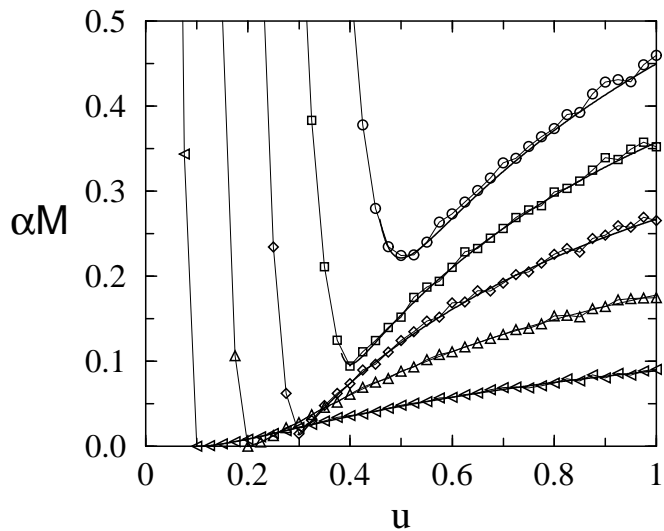}\\
\end{center}
\caption{Overlap parameter $\alpha M$ vs $u$ in the fully connected model ($c=1$) for different values of $\alpha$. Solid lines are the analytical predictions (\ref{eq:emm}) for the ergodic phase $u>u^\star(\alpha,c,\Gamma)$. Connected markers represent results from simulations (circles: $\alpha=1$, squares: $\alpha=0.8$, diamonds: $\alpha=0.6$, triangle up: $\alpha=0.4$, left triangle: $\alpha=0.2$). Simulation parameters as in Fig. \ref{fig:qphi}.}
\label{fig:emm}
\end{figure}

It is in general very tedious to measure response functions directly
in simulations. To confirm the predicted divergence of $\chi$ in the
fully connected model $c=1$, we therefore note that in the course of
the generating functional calculation the covariance matrix
$\Lambda(t,t')=\avg{\eta(t)\eta(t')}_\star/\alpha=[(\id-\bG)^{-1}\bC(\id-\bG^T)^{-1}]_{tt'}$
of the effective-species noise also turns out the indicate the
temporal correlations of the following order parameters
\footnote{Note that the parameter $m^\mu(t)=N^{-1/2}\sum_i\xi_i^\mu
x_i(t)$ is closely related to what is known as an `overlap parameter'
in the context of neural networks.}:
\be
m^\mu(t)\equiv\frac{1}{\sqrt{N}}\sum_{i=1}^N\xi_i^\mu x_i(t).
\ee
More precisely one has
\be
\Lambda(t,t')=\frac{1}{\alpha N}\sum_{\mu=1}^{\alpha N}\avg{m^\mu(t)m^\mu(t')}_*.
\ee
At the fixed point in the ergodic regime we therefore have (with $m^\mu=\lim_{t\to\infty}m^\mu(t)$)
\be\label{eq:emm}
M\equiv\frac{1}{\alpha N}\sum_{\mu=1}^{\alpha N}\avg{(m^\mu)^2}_\star=\frac{q}{(1-\chi)^2}.
\ee
In particular we expect $M\to 0$ as $\chi\to -\infty$ at $u=\alpha/2$ for $\alpha<1/2$. This is indeed confirmed in simulations, see Fig. \ref{fig:emm}. No zeroes of $M$ are observed for any value of $u$ when $\alpha>1/2$.

\subsection{Instability of the fixed point}
In this section we will discuss the local
stability of the fixed point reached by the system. To this end we
will follow \cite{OD} and take into account small fluctuations $y(t)$
about the fixed point $x$ attained by the effective species, i.e. we
will write $x(t)=x+\varepsilon y(t)$ in Eq.  (\ref{eq:effprocess}),
and will accordingly also assume that the single-particle noise is of
the form $\eta(t)=\eta+\varepsilon v(t)$, with $v(t)$ a
(time-dependent) fluctuation about the fixed point value $\eta$. The
magnitude of these fluctuations $\varepsilon$ is taken to be small. We
will also allow for additional white noise $\varepsilon\zeta(t)$ of a
small amplitude $\varepsilon$ (with
$\avg{\zeta(t)\zeta(t')}=\delta(t-t')$), and will then study the
stability of the fixed point $(x,\eta)$ with respect to these
fluctuations. Note that self-consistency (cf. Eq. (\ref{eq:noise}))
dictates that \be\label{eq:vv}
\avg{v(t)v(t')}=\alpha[c(\id-\bG)^{-1}\bD(\id-\bG^T)^{-1}+(1-c)\bD](t,t'), \ee
where we write $D(t,t')=\avg{y(t)y(t')}_\star$.

We first note that these fluctuations decay exponentially for species which are dying out asymptotically, i.e. for whom the fixed point is given by $x=0$. In this case insertion of $x(t)=\varepsilon y(t)$ and $\eta(t)=\Sigma z+\varepsilon v(t)$ into (\ref{eq:effprocess}) gives
(to linear order in $\varepsilon$)
\be\label{eq:zstable}
\frac{d}{dt}y(t)=-y(t)\left(\Sigma z-\lambda\right),
\ee
and we have $\Sigma z-\lambda>0$ for fixed points $x(z)=0$ as seen in (\ref{eq:xofz}).

For fixed points $x=x(z)>0$ we find instead (again to linear order in $\varepsilon$)
\be
\fl~~~\frac{d}{dt}y(t)=-x\left(\frac{2u}{c}y (t)+\alpha\int_{0}^t d\tp [c\bG(\id-\bG)^{-1}+\Gamma(1-c)\bG](t,\tp)y(\tp)+v(t)+\zeta(t)\right),
\ee
which is conveniently expressed in terms of Fourier transforms $\{\widetilde y(\omega),\widetilde v(\omega),\widetilde \zeta(\omega),\widetilde G(\omega)\}$ of $\{y(t),v(t),\zeta(t),G(t)\}$ as 
\be
\widetilde y(\omega)=-\frac{\widetilde v(\omega)+\widetilde\zeta(\omega)}{i\omega/x+2u/c+\alpha c\widetilde G(\omega)/(1-\widetilde G(\omega))+\alpha(1-c)\Gamma\widetilde G(\omega)}.
\ee
 Focusing on $\omega=0$ and using $\avg{\zeta(t)\zeta(t')}_\star=\delta(t-t')$ as well as
\be
\avg{|\widetilde v(0)|^2}_\star=\alpha\avg{|\widetilde y(0)|^2}_\star\left(\frac{c}{(1-\chi)^2}+(1-c)\right)
\ee
(which follows from (\ref{eq:vv})) we have
\be
\widetilde D(0)\equiv \avg{|\widetilde y(\omega=0)|^2}_\star=\frac{\phi\left(\alpha\widetilde D(0)\left[c(1-\chi)^{-2}+(1-c)\right]+1\right)}{[2u/c+\alpha c\chi/(1-\chi)+\alpha(1-c)\Gamma\chi]^2},
\ee
where $\widetilde D(\omega)$ is the Fourier transform of the correlation function  $D(\tau)=\avg{y(t+\tau)y(t)}_\star$ in the stationary state. The factor $\phi$ (the fraction of surviving species) takes into account the earlier result that perturbations $y(t)$ about fixed points $x=0$ do not contribute to the long-time behaviour of the correlation function $D$. We then find
\be\label{eq:dtilde}
\fl ~~\widetilde D(0)=\left[\phi^{-1}\left(2u/c+\alpha c\frac{\chi}{1-\chi}+\alpha\Gamma(1-c)\chi\right)^2-\alpha\left(c(1-\chi)^{-2}+(1-c)\right) \right]^{-1},
\ee
so that $\widetilde D(\omega=0)=\avg{|\widetilde
y(\omega=0)|^2}_\star$ diverges when the square bracket on the
right-hand-side becomes zero. This conditions defines a transition
point $u=u^\star(\alpha,c,\Gamma)$ and the divergence of $\widetilde D(\omega=0)$
at $u^\star$ suggests that the assumed fixed points become
unstable. $\widetilde D(\omega=0)$ is predicted to become negative
below the transition, when $\left(2u/c+\alpha c\frac{\chi}{1-\chi}+\Gamma(1-c)\chi\right)<\phi\alpha\left(c(1-\chi)^{-2}+(1-c)\right)$,
leading to a further contradiction and indicating that our theory
cannot be continued below $u^\star$ \footnote{Further details regarding
this type of transition can also be found in \cite{OD,OD2}:
Instabilities of the above type have been associated with $1/\omega$
(or similar) divergencies of $D(\omega)$ at the transition point
$u=u^\star(\alpha,c,\Gamma)$ as $\omega\to 0$ in Gaussian replicator models,
indicating that fluctuations $y(t)$ are no longer damped and the fixed
points become unstable at $u=u^\star(\alpha,c,\Gamma)$. Above the transition
$\avg{y(t+\tau)y(t)}_\star$ decays as $1/\tau^2$ as $\tau\to\infty$,
suggesting that the fixed points are local attractors of the
dynamics.}. Using saddle point equation (\ref{eq:sp3}) this onset of
instability occurs when
\be\label{eq:instability}
\frac{\alpha}{\phi}\left(c\frac{\chi^2}{(1-\chi)^2}+(1-c)\chi^2\right)=1.
\ee
For $c=1$ this relation is identical to the condition marking the
onset of a de Almeida-Thouless (AT) instability of the replica
symmetric solution identified in the analysis of the statics of the
fully connected model \cite{SF}.

\subsection{Breakdown of weak-long term memory}
While in the previous section we have related the breakdown of our
ergodic theory to a local instability of the fixed point reached by
the dynamics, we will now inspect for an onset of long-term memory at
finite integrated response. This type of transition has been observed
previously for example in Minority Games with self-impact correction
or diluted interactions \cite{corr,dilute}, and can also be
interpreted in terms of a breakdown of time-translation invariance, see
\cite{Book2} for details. In order to see how solutions with memory bifurcate from the time-translation invariant ergodic states we will make the
following ansatz for the response function
\be\label{eq:g0ghat}
G(t,t')=G_0(t-t')+\varepsilon\widehat G(t,t'),
\ee
where $\varepsilon\widehat G(t,t')$ is a small contribution which breaks
time-translation invariance. In the context of the present replicator
model the time-translation invariant part $\bGz$ reflects the
contribution of the surviving species to the total response function
as discussed above. Note that asymptotically extinct species do not
contribute to the response in time-translation invariant states, as
fixed points $x_i=0$ are insensitive to small perturbations as seen in
the previous section. Perturbations applied during the transients of
the dynamics, however, can leave their traces in the choice of the
particular overall asymptotic fixed point $\bx=(x_1,\dots,x_N)$: the
specific set of of surviving species and their stationary
concentrations may well depend on those early perturbations. This
effect is accounted for by the function $\widehat G(t,t')=\widehat
G(t')$, which is taken to depend only on the time $t'$ at which the
perturbation is applied, but not on the later time $t$ at which the
effect of the perturbation is measured \cite{Book2}.

Starting from (\ref{eq:g0ghat}) one expands the kernel of the retarded self-interaction in the effective process to linear order in $\widehat G$, and finds
\be
\fl~~~c\bG(1-\bG)^{-1}+\Gamma(1-c)\bG =c\bGz(1-\bGz)^{-1}+\Gamma(1-c)\bGz+\varepsilon\bR+{\cal O}(\varepsilon^2),
\ee
where 
\be
R(t,t')=\Gamma(1-c)\widehat G(t,t')+c\sum_{n=0}^\infty\sum_{m=0}^{n-1}\left[(\bGz)^m\widehat\bG(\bGz)^{n-m-1}\right](t,t').
\ee
Taking into account this extra contribution we find that fixed points are now to be determined as (with $\chi=\sum_t G_0(t)$)
\be
x=\frac{\lambda-\eta-\varepsilon\alpha\int_0^{t}dt'R(t,t')x(t')}{2u/c+\alpha c\chi/(1-\chi)+\alpha\Gamma(1-c)\chi}\theta\left[\lambda-\eta-\varepsilon\alpha\int_0^{t}dt'R(t,t')x(t')\right].
\ee
Taking the derivative with respect to a perturbation at time $t''$ followed by averaging we then have (to first order in $\varepsilon$)
\be
\fl~~~\widehat G(t'')=-\frac{\alpha}{2u/c+\alpha c \chi/(1-\chi)+\alpha\Gamma(1-c)\chi}\int_0^{t}dt'R(t,t')\avg{\theta\left[\lambda-\eta\right]\frac{\delta x(t')}{\delta h(t'')}}_\star.
\ee
Within our ansatz the average on the right-hand-side is identified as $G_0(t',t'')$. Using the above expression for $\bR$ and introducing the shorthand $\widehat \chi=\sum_t \widehat G(t)$ we then find (after suitable integration over $t''$ and averaging over $t$)
\be\label{bloe}
\fl~~~\widehat \chi=-\frac{\alpha}{2u/c+\alpha c \chi/(1-\chi)+\alpha\Gamma(1-c)\chi}\left(c \frac{\chi}{(1-\chi)^2}+\Gamma (1-c)\chi\right)\widehat\chi.
\ee
Although $\widehat\chi=0$ is always a solution, non-zero branches may
bifurcate from the time-translation invariant solution at the point at
which the coefficient in front of $\widehat\chi$ on the
right-hand-side of (\ref{bloe}) becomes one. Using Eq. (\ref{eq:sp3})
this condition is given by

\begin{figure}
\vspace*{1mm}
\begin{center}
\epsfxsize=80mm  \epsffile{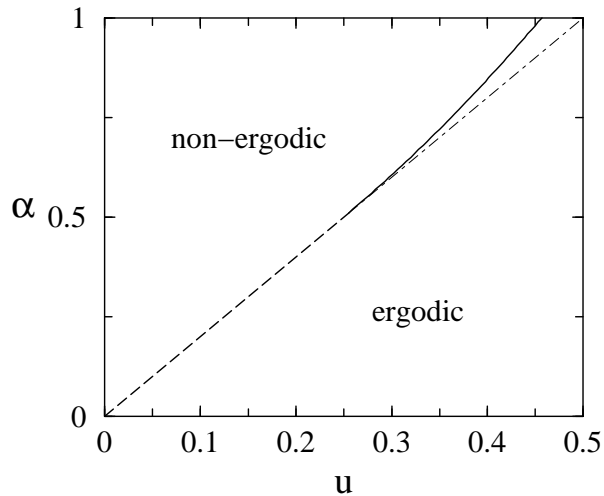}\\
\end{center}
\vspace*{-3mm}
\caption{Phase diagram of the model. The dashed line $u=\alpha/2$ for $\alpha<1/2$ marks the discontinuous transition signalled by a divergence of the integrated response. The solid line for $\alpha>1/2$ corresponds to the simultaneous onset of long-term memory and of an instability of the fixed point against small perturbations. The line $u=\alpha/2$ has been continued as a dot-dashed line for $u>\alpha/2$. }
\label{fig:phasediagram}
\end{figure}
\be\label{eq:mo}
\frac{\alpha}{\phi}\left(c \frac{\chi^2}{(1-\chi)^2}+\Gamma (1-c)\chi^2\right)=1.
\ee
For fully symmetric dilution ($\Gamma=1$) this is identical to the
above condition (\ref{eq:instability}) marking the onset of the
instability of the fixed points and for $c=1$ furthermore coincides with the
condition signalling the AT-instability of the replica-symmetric
solutions of the statics of the fully connected model \cite{O,SF}. For $\Gamma<1$ (and $c<1$) one observes that a fulfillment of memory-onset condition (\ref{eq:mo}) implies that the instability of the fixed points has already set in (as signalled by a negative right-hand-side of (\ref{eq:dtilde})). For asymmetrically diluted couplings the onset of instability thus occurs first as $u$ is lowered at fixed $\alpha,c,\Gamma$ and the MO-line (\ref{eq:mo}) has no physical meaning. A physical interpretation of the two types of transitions in the case of (partially) asymmetric dilution will be discussed briefly below.

\subsection{Phase diagram of the fully connected model}
The resulting phase diagram of the fully connected model ($c=1$) in
the $(u,\alpha)$-plane is shown in Fig. \ref{fig:phasediagram} and
coincides with the one found in the statics \cite{O,SF}. For values of
$\alpha$ greater than $1/2$ we do not find any divergence of any of
the order parameters, but rather a continuous onset of long-term
memory occurs at a point $u^\star=u^\star(\alpha)$ slightly above the line
$u=\alpha/2$, accompanied by a simultaneous onset of an instability of
the assumed fixed point against small fluctuations. All order
parameters are continuous across this transition.

For $\alpha<1/2$ one finds a divergence $\chi\to -\infty$ at
$u^\star(\alpha)=\alpha/2$ as discussed above. This transition is marked
by discontinuities in the order parameters such as $q$ and $\phi$, see
Fig. \ref{fig:qphi}. We also note that the overlap parameter $M$
vanishes at this point $u=u^\star(\alpha)$ for $\alpha<1/2$, but that
simulations suggest that it remains strictly positive below this
transition ($u<u^\star(\alpha)$) as indicated in Fig. \ref{fig:emm}. This
is at variance with the behaviour of the analogous quantity $H$ in
Minority Games \cite{Book1, Book2}, which is positive above the
corresponding transition of the Minority Game, vanishes at the point
at which $\chi$ diverges and then remains identically zero below the
transition point\footnote{The so-called `predictability' $H$ studied
in Minority Games is analogous to the overlap parameter $M$ discussed
here in the sense that both quantities are proportional to the
covariance of the effective single-particle noises of the respective
models at infinite time-separation.}. In simulations (not shown here)
we also do not observe any discontinuities of $q$ or $\phi$ as
$\alpha$ is varied at constant $u$ strictly within in the non-ergodic
phase (i.e. not crossing the line segment
$u=\alpha/2,\,0\leq\alpha\leq 1/2$). In this sense the non-ergodic
phase indicated in Fig. \ref{fig:phasediagram} does not appear to be
divided into two separate regions in an obvious way (for example one
with $\chi=-\infty, M=0$ and another with $\chi$ finite and
$M>0$). 

As shown in Figs. \ref{fig:qphi} and in the left panel of
Fig. \ref{fig:distance} the analytically obtained location
$u^\star(\alpha,c=1)$ of the phase transition in the fully connected
model agrees (within the accuracy of our simulations) with the
onset of deviations from the ergodic theory and of the sensitivity to
initial conditions
\footnote{Note that $d^2$ as depicted in Fig. \ref{fig:distance} is a
measure for the distance between the microscopic fixed points reached
from different realisations of the random initial conditions. In
contrast to Minority Games where the so-called `market volatility'
crucially depends on the starting point in the non-ergodic regime of
the game we have not been able to identify an observable which can be
expressed in terms of the dynamic order parameters $\{\bC,
\bG\}$ (and possibly $\blambda$) and which shows a dependence on the
type of initial conditions below $u^\star$.}\footnote{Note also that in \cite{corr} {\em small} perturbations were applied in a similar situation in the context of the Minority Game to verify the onset of memory. We have tried to measure the response of the present system to small perturbations at early and late times respectively, but found the resulting simulations to be inconclusive. The data in Fig. \ref{fig:distance} is therefore restricted to separate starts from two independent sets of initial conditions at fixed disorder.}.

The general picture that emerges here seems to indicate that
in the ergodic regime $u>u^\star(\alpha,c=1)$ only one fixed point is found,
so that the microscopic dynamics ends up in the same configuration
independently of initial conditions. Below $u^\star(\alpha,c=1)$ all
trajectories still evolve into microscopic fixed points, but these do
not seem to be unique, but possibly of an exponential (in $N$) 
number \cite{DO,OD}, leading to the identified dynamic instabilities and
the breaking of ergodicity (and to a breakdown of the
replica-symmetric solution in the statics). Further studies of the
non-ergodic regime, possibly along the lines of
\cite{DO} might however be appropriate to understand the nature of this
phase in more detail. Given the memory effects below $u^\star$ an analysis
of the dynamics below the transition is very likely to require a
solution in terms of the full two-time quantities $C(t,t')$ and
$G(t,t')$ during the transients of the system's temporal evolution.

\subsection{Effects of dilution}

We will now turn to the consequences of diluting the Hebbian couplings. A
brief discussion of these effects can also be found in \cite{SF},
albeit based on replica methods and hence restricted to the case of
symmetric dilution. 

First note that the relations (\ref{eq:instability}) and (\ref{eq:mo})
signalling the instability of the fixed points and the onset of memory
respectively coincide for symmetric dilution $\Gamma=1$. In this case
a Lyapunov function of the replicator dynamics can be found and the
system evolves into a fixed point irrespective of
$\{u,\alpha,c\}$. Below $u^\star(\alpha,c,\Gamma=1)$ exponentially
many fixed points exist and initial conditions as well as
perturbations in the transients determine which of these is reached by
the dynamics, resulting in broken ergodicity and memory effects. For
$\Gamma<1$ the instability condition (\ref{eq:instability}) is
fulfilled first as $u$ is lowered from large values (at constant
$\alpha,c,\Gamma<1$). While above $u^\star$ one single stable fixed
point is found, the number of fixed points is expected to be
exponentially (in $N$) suppressed for asymmetric couplings below the
transition \cite{OD,OD2}. Since no phase with multiple fixed points is
found for $\Gamma<1$ no MO-transition occurs (the identification of
which was based on a fixed-point assumption also below $u^\star$) and
the line in the phase diagram defined by (\ref{eq:mo}) has no physical
meaning as it is preceded by condition (\ref{eq:instability}). The
relation between the MO- and instability condition is discussed in
more detail also in
\cite{myother}. Furthermore no singularities $\chi\to -\infty$ are
found for $c<1$ before the instability condition sets in, resulting in
the location of all transitions being determined by
(\ref{eq:instability}) in this case.

\begin{figure}[t]
\vspace{2cm}
\hspace*{35mm} \setlength{\unitlength}{1.1mm}
\begin{tabular}{cc}
\begin{picture}(100,55)
\put(-26,5){\epsfysize=53\unitlength\epsfbox{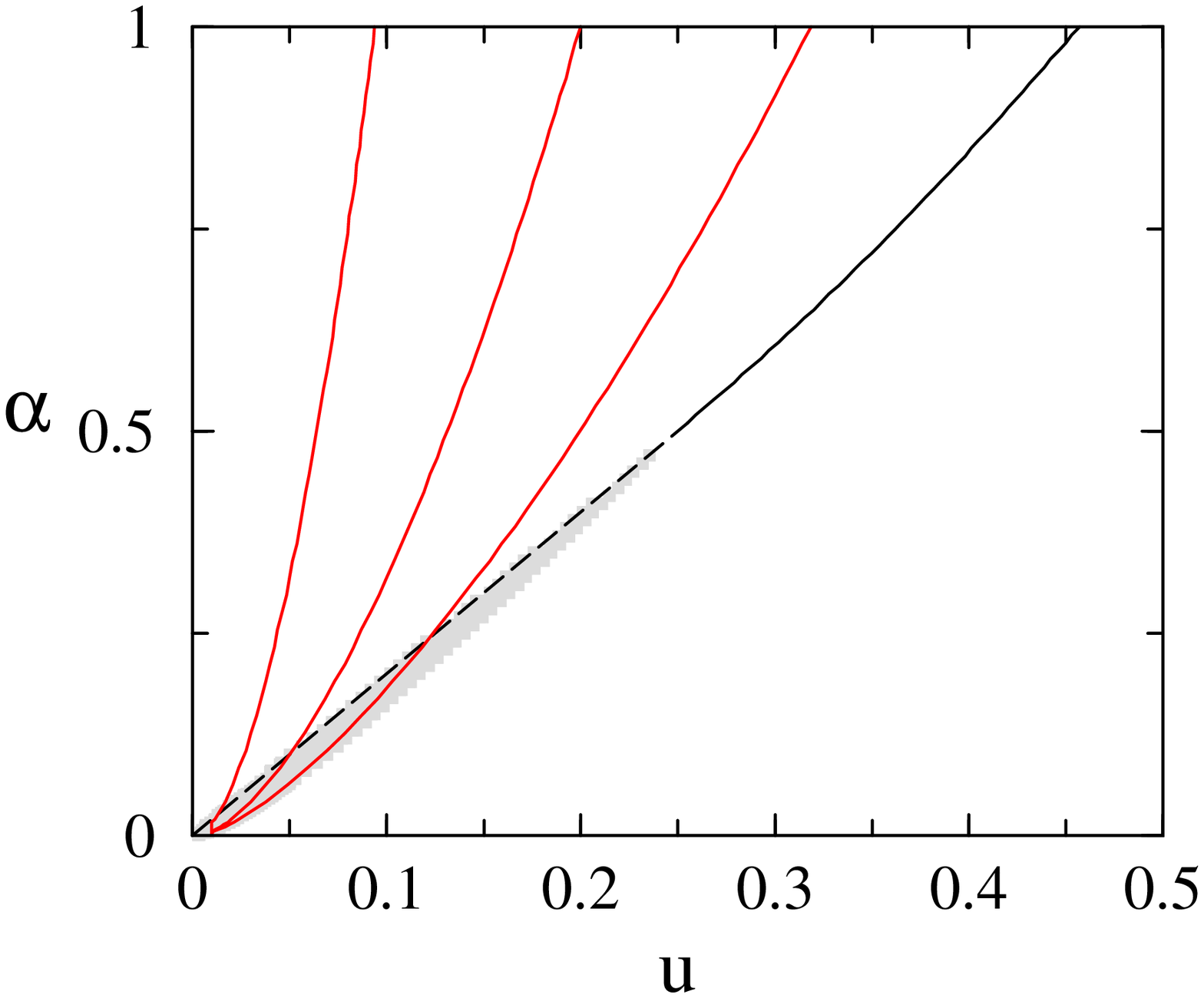}}

\put(-18,67){\footnotesize $c=0.25$}
\put(1,67){\footnotesize $c=0.5$}
\put(20,67){\footnotesize $c=0.75$}
\put(-14,65){\vector(1,-1){6}}
\put(4,65){\vector(0,-1){6}}
\put(24,65){\vector(-1,-1){6}}
\put(68,67){\footnotesize $c=0.25$}
\put(84,67){\footnotesize $c=0.5$}
\put(96,67){\footnotesize $c=0.75$}
\put(73,65){\vector(0,-1){6}}
\put(87,65){\vector(0,-1){6}}
\put(101,65){\vector(-1,-2){3}}
\end{picture} &
\begin{picture}(100,55)
\put(-57,5){\epsfysize=53\unitlength\epsfbox{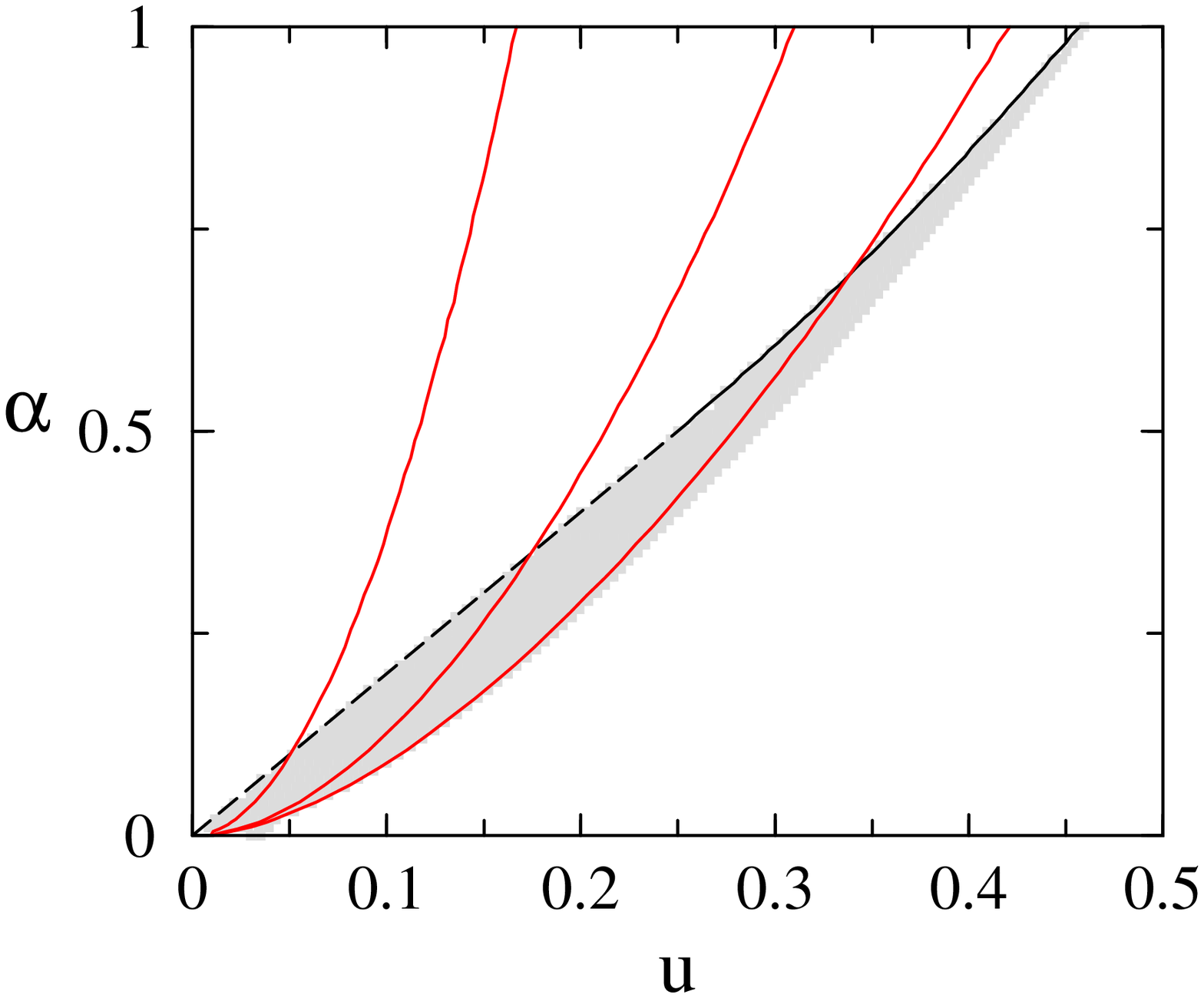}}
\put(-15,19){$\Gamma=1$}
\put(-85,19){$\Gamma=0$}
\end{picture}
\end{tabular}

\vspace*{4mm} \caption{Phase diagrams of the models with fully asymmetric and with fully symmetric dilution. The curves in the $(u,\alpha)$-plane mark the onset of memory at constant $c=0.25,~0.5,~0.75$. Ergodic phases are found to the bottom-right of the curves. The transition line of the fully connected model $c=1$ is shown for comparison. The grey areas are sketches of the re-entrance regions, here the system is ergodic for small $c\ll 1$, non-ergodic in an intermediate regime and ergodic again at large $c$ close to $c=1$.}
\label{fig:pgdil}
\end{figure}
The phase diagrams for fully symmetric and fully asymmetric dilution
are shown in Fig. \ref{fig:pgdil}. With the present scaling of the
self-interactions (with $c$) the general effect of increasing dilution
appears to be an increase the range of ergodicity, only in a region
directly below the transition line of the fully connected model do we
find re-entrance behaviour: here the system is ergodic at $c>c_2$ and
fixed $(u,\alpha)$, becomes non-ergodic at in an intermediate interval
$c\in[c_1,c_2]$ and then ergodic again for $c<c_1$. We have performed
some simulations to confirm these findings (not shown here). While the
numerical experiments are consistent with re-entrance behaviour the
exact location of the window of non-ergodicity is hard to verify
numerically especially because $c_2$ is found to lie close to $c=1$ in
the analytical solution.

\begin{figure}[t]
\vspace*{1mm}
\begin{tabular}{cc}
\epsfxsize=72mm  \epsffile{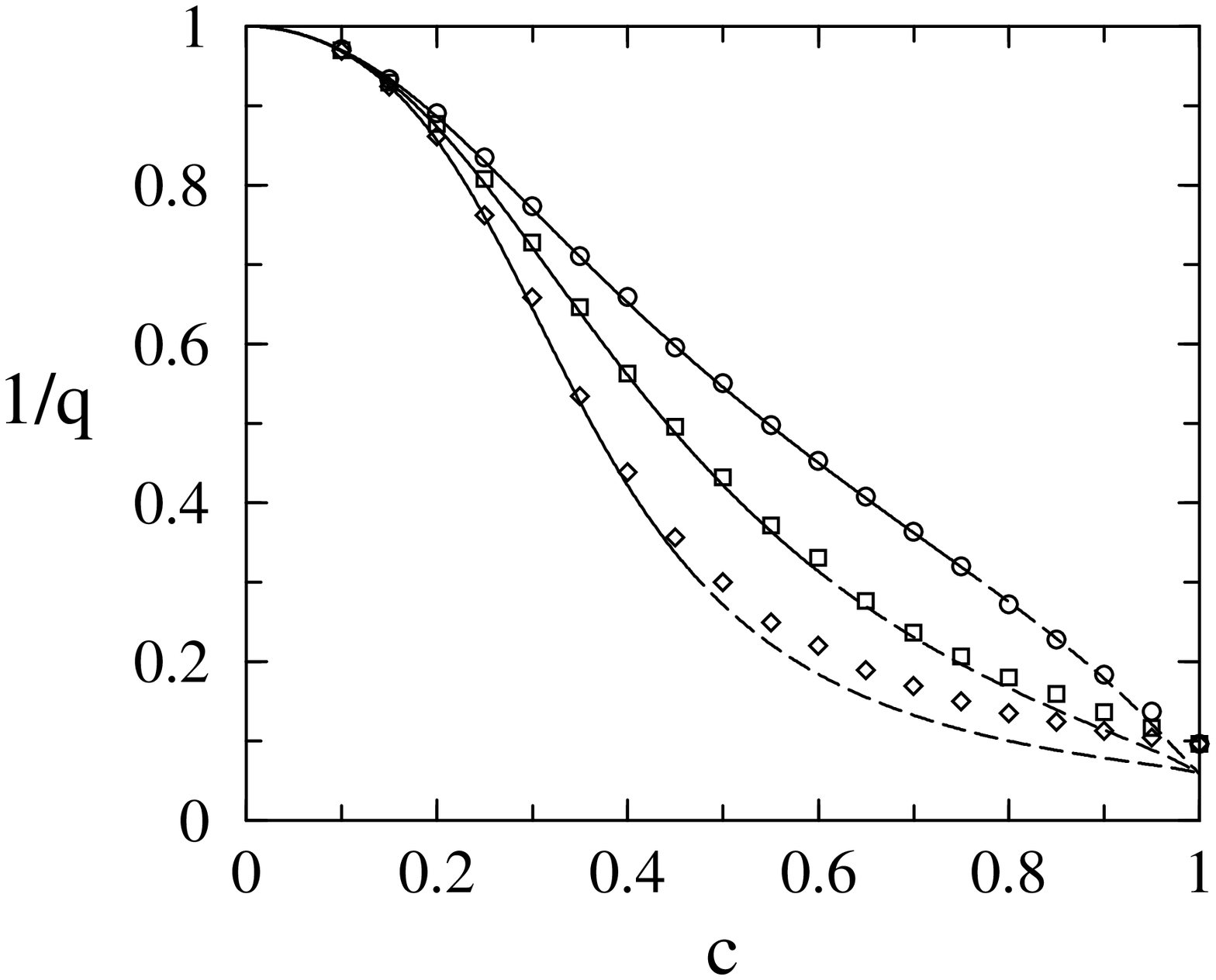} ~&~~
\epsfxsize=72mm  \epsffile{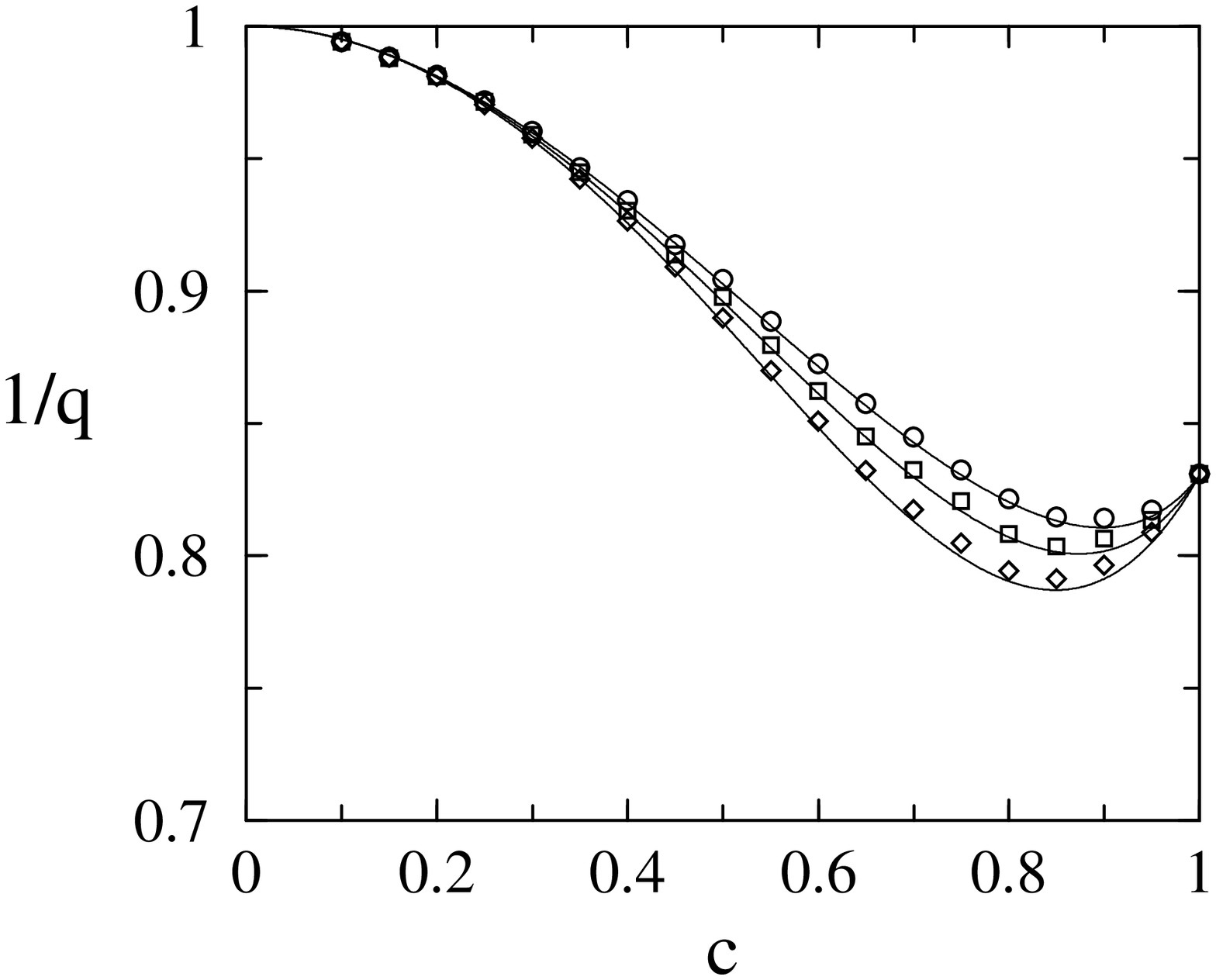}
\end{tabular}
\vspace*{4mm} \caption{Inverse order parameter $1/q$ at fixed $(u,\alpha,\Gamma)$ as a function of the connectivity for the dilute system. Left: $u=0.2, \alpha=0.5$, right: $u=0.5,\alpha=0.5$. Solid lines are obtained from the ergodic theory and have been continued as dashed lines into the non-ergodic phase. Symbols are from simulations with $N=500$ species, run for $20000$ discretisation steps, averaged over $50$ samples of the disorder (circles: $\Gamma=0$, squares: $\Gamma=0.5$, diamonds: $\Gamma=1$). }
\label{fig:qdil}
\end{figure}

Finally we plot the reciprocal order parameter $1/q$ versus the
connectivity $c$ at fixed $(u,\alpha,\Gamma)$ in
Fig. \ref{fig:qdil}. While the scaling of the self-interaction used
here, both purely decreasing behaviour with increasing connectivity as
well as extremal diversity at intermediate $c$ can be found depending
on the values of $u$ and $\alpha$. Similar non-monotonic behaviour is
also reported in \cite{SF} for the case of symmetric dilution,
although we note here that adapting our theory to a model with
$J_{ii}c_{ii}/c={\cal O}(c^0)$ suggest that this non-monotonicity and
the above re-entrance effects appear are to be absent if the
self-interaction does not scale with $c$ (results not shown). In
general one observes that the diversity $1/q$ (and the fraction of
surviving species) is higher for asymmetric than for symmetric
dilution.

\section{Finite number of traits}
We now turn to the case of a finite number $p$ of traits. In this case
generating functionals are not required to formulate a closed
dynamical theory, instead it is sufficient to study the evolution of the
probability distributions of species concentrations on so-called
sub-lattices, in which all species with the same trait vector
$\bxi_i=(\xi_i^{\mu=1},\dots,\xi_i^{\mu=p})$ are treated
collectively. We here concentrate on the fully connected model.

The replicator equations take the form
\be
\frac{d}{dt}x_i(t)=-x_i(t)\left(\sum_j J_{ij}x_j(t)-\frac{1}{N}\sum_{jk}J_{jk}x_j(t)x_k(t)\right)
\ee
with 
\be
J_{ij}=N^{-1}\sum_{\mu=1}^p\xi_i^\mu\xi_j^\mu,~ i\neq j,~~~~~J_{ii}=2u.
\ee
It is known from the statics that a model with a finite number of traits chosen with equal probability from $\xi_i^\mu\in\{-1,1\}$ leads to a trivial stationary state (with fixed points $x_i\equiv 1$ for all $i$) \cite{OF3} so that we follow \cite{OF3} and introduce a bias parameter $a\in[0,1]$ and draw the traits independently from a distribution
\be
P(\xi_i^\mu)=\frac{1+a}{2}\delta_{\xi_i^\mu,1}+\frac{1-a}{2}\delta_{\xi_i^\mu,-1}
\ee
for all $i$ and $\mu$.

Analytical progress can here be made without the use of generating
functionals. It is sufficient to introduce sub-lattices
\be
I_{\bta}=\{i:\bxi_i=\bta\}\subset\{1,\dots,N\}
\ee
with $\bta\in\{-1,1\}^p$ and to consider the macroscopic observables
\be
\rho_{\bta}(x,t)=\avg{|I_{\bta}|^{-1}\sum_{i\in I_{\bta}}\delta(x-x_i(t))},
\ee
in terms of which closed dynamical equations can be formulated along
the lines of \cite{Cool00b}. $\avg{\dots}$ here denotes an average
with respect to the time-dependent microscopic state density
$p_t(x_1,\dots,x_N)$. With the relative proportions
$p_{\bta}=\lim_{N\to\infty} |I_{\bta}|/N$ the Fokker-Planck equation
governing the evolution of $p_t(x_1,\dots,x_N)$ can be reduced to the
following set of $2^p$ equations closed in the $2^p$ probability
distributions $\{\rho_{\bta}(x,t):\bta\in\{-1,1\}^p\}$, valid in the
thermodynamic limit $N\to\infty$:
\BE
\frac{d}{dt}\rho_{\bta}(x)&=&-\frac{\partial}{\partial x}\left[x\left(-2ux-\sum_{\btap}p_{\btap}~\bta\cdot\btap\int dx' \rho_{\btap}(x')x'\right.\right.\\
&&\left.\left.+\sum_{\btap,\btapp}p_{\btap}p_{\btapp}~\btap\cdot\btapp\int{dx'}\rho_{\btap}(x')x'\int dx'' \rho_{\btapp}(x'')x''\right.\right.\\
&&\left.\left.+2u\sum_{\btap} p_{\btap}\int dx' \rho_{\btap}(x') x'^2 \right)\right].
\EE
We have here suppressed the time-dependence of the densities $\rho_{\bta}(x)$ and use the shorthand $\bta\cdot\btap=\sum_{\mu=1}^p\eta_\mu\eta'_\mu$. Upon the introduction of the moments 
\be
\nu^{(k)}_{\bta}(t)=\int_0^\infty dx~ \rho_{\bta}(x)x^k
\ee
($k=0,1,2,\dots$) one has
\BE
\frac{d}{dt}\rho_{\bta}(x)&=&-\frac{\partial}{\partial x}\left[x\left(-2ux-\sum_{\btap}p_{\btap}~\bta\cdot\btap\nu^{(1)}_{\btap}\right.\right.\nonumber\\
&&\left.\left.+\sum_{\btap,\btapp}p_{\btap}p_{\btapp}~ \btap\cdot\btapp\nu^{(1)}_{\btap}\nu^{(1)}_{\btapp}+2u\sum_{\btap} p_{\btap}\nu^{(2)}_{\btap}\right)\right]\label{eq:tata}.
\EE
Due to the effective non-linearity (in $x$) on the right-hand-side a
simple solution in terms for example of an Ornstein-Uhlenbeck process
\cite{Cool00b} appears impossible.  Multiplication of (\ref{eq:tata}) by $x^k$ on both
sides and subsequent integration over $x$ then leads to
\be
\frac{d}{dt}\nu^{(k)}_{\bta}=-2ku\nu^{(k+1)}_{\bta}-k\nu^{(k)}_{\bta}\left[\sum_\mu\eta_\mu m_\mu-\sum_\mu m_\mu m_\mu-2u\sum_{\btap} p_{\btap}\nu^{(2)}_{\btap}\right],\label{eq:hir}
\ee
with $m_\mu=\sum_{\bta}p_{\bta}\eta_\mu\nu^{(1)}_{\bta}$. Eqs. (\ref{eq:hir}) form a closed, but hierarchical system of equations for the time-evolution of the moments $\{\nu_{\bta}^{(k)}(t)\}$.

We will now assume symmetrical initial conditions $m_\mu(t=0)\equiv m(0)$ for all $\mu$, and will accordingly write $m_\mu(t)\equiv m(t)$ for all $\mu$ for all times $t$. Furthermore the densities $\rho_{\bta}(x)$ will depend on the sub-lattice structure only via the number of entries $+1$ in the vector $\bta$, i.e. all indices $\bta$ may effectively be replaced by $n(\bta)=|\{\mu:\eta_\mu=1\}|$, and the $p_{\bta}$ translate into probabilities
\be
W_n=\left(\begin{array}{c}n\\p\end{array}\right)\left(\frac{1+a}{2}\right)^n\left(\frac{1-a}{2}\right)^{p-n},~~n=0,1,2,\dots,p
\ee 
for the occurance of a trait vector $\bta$ with $n$ entries equal to $+1$ (and $p-n$ entries equal to $-1$). The overlap can then we written as
\be\label{eq:moftf}
m(t)=\sum_{n=0}^p W_n \nu^{(1)}_n(t) \left(\frac{2n}{p}-1\right)
\ee
(where we have used $\lim_{N\to\infty}|I_{\bta}|^{-1}\sum_{i\in I_{\bta}}\eta_\mu=(2n(\bta)/p)-1~\forall\mu$). The dynamical equations for the first two moments simplify to
\BE
\frac{d}{dt}\nu^{(1)}_n&=&-2u\nu^{(2)}_{n}-\nu^{(1)}_{n}\left[mp\left(\frac{2n}{p}-1\right)-p m^2-2u\sum_{n'=0}^p W_{n'}\nu^{(2)}_{n'}\right]\\
\frac{d}{dt}\nu^{(2)}_n&=&-4u\nu^{(3)}_{n}-2\nu^{(2)}_{n}\left[mp\left(\frac{2n}{p}-1\right)-p m^2-2u\sum_{n'=0}^pW_{n'}\nu^{(2)}_{n'}\right].
\EE
Note that this set of equations does not close as the evolution of the second moments couples to the third moments, the evaluation of which requires the fourth moments and so on. In general one has
\BE
\fl ~~~~~\frac{d}{dt}\nu^{(k)}_n&=&-2ku\nu^{(k+1)}_{n}-k\nu^{(k)}_{n}\left[mp\left(\frac{2n}{p}-1\right)-p m^2-2u\sum_{n'=0}^pW_{n'}\nu^{(2)}_{n'}\right].
\EE
A brief remark on the stationary state are is in order at this point. Indeed it is straightforward to verify that the static solution $\rho_n(x)=\delta(x-x_n)$ with
\be
x_n=1+\frac{pa(a+1)-2an}{2u+1-a^2}, ~~n=0,1,\dots,p \label{fpp}
\ee
 found in \cite{OF3} and resulting in $m=2ua/(2u+1-a^2)$ is a fixed
 point of the above dynamical equations. We here assume that
\be\label{eq:condition}
p<1+\frac{1}{a}+\frac{2u}{a(1-a)}
\ee
so that all species survive in the long-term limit. A more detailed
discussion regarding the fraction of surviving species whenever this
condition is not fulfilled can be found in \cite{OF3}.

In order to proceed with the analysis of the dynamics an approximation is
required at this stage, effectively truncating the above hierarchical
set of equations. We here restrict the further discussion to the most
simple approximative scheme and take the densities $\rho_n(x)$ to be
Gaussian. This allows to express each third moment $\nu_n^{(3)}$ in
terms of $\nu_n^{(1)}$ and $\nu_n^{(2)}$ and we find the following
approximate but closed set of $2(p+1)$ equations for the
$\{\nu^{(1)}_n,\nu^{(2)}_n\}$, $n=0,1,\dots,p$:
\BE
\fl \frac{d}{dt}\nu^{(1)}_n&=&-2u\nu^{(2)}_{n}-\nu^{(1)}_{n}\left[mp\left(\frac{2n}{p}-1\right)-p m^2-2u\sum_{n'}W_{n'}\nu^{(2)}_{n'}\right]\label{eq:lu}\\
\fl\frac{d}{dt}\nu^{(2)}_n&=&-4u[\nu^{(1)}_n(3\nu^{(2)}_n-2(\nu^{(1)}_n)^2)]-2\nu^{(2)}_{n}\left[mp\left(\frac{2n}{p}-1\right)-p m^2-2u\sum_{n'}W_{n'}\nu^{(2)}_{n'}\right]\label{eq:la}.
\EE
We will write $\sigma_n^2\equiv \nu_n^{(2)}-(\nu_n^{(1)})^2$ in the
following. The Gaussian approximation can be expected to be accurate at
most in a regime where the $\rho_n(x)$ are highly peaked around
positive means, i.e. in which $\nu_n^{(1)}>0$ and
$\sigma_n^2/(\nu_n^{(1)})^2\ll 1$. In particular any Gaussian
approximation with $\sigma_n^2>0$ is inconsistent in the sense that it
assigns finite probability to negative species concentrations, which are
excluded {\em a priori} by the replicator equations. Note however that our
approximative ansatz consists only of assuming the relations between
the first three moments of Gaussian distributions to hold for the
densities $\rho_n(x)$. No other properties of Gaussian distributions
are used here.

\begin{figure}[t]

\hspace*{35mm} \setlength{\unitlength}{1.1mm}
\begin{tabular}{cc}
\begin{picture}(100,55)
\put(-26,5){\epsfysize=48\unitlength\epsfbox{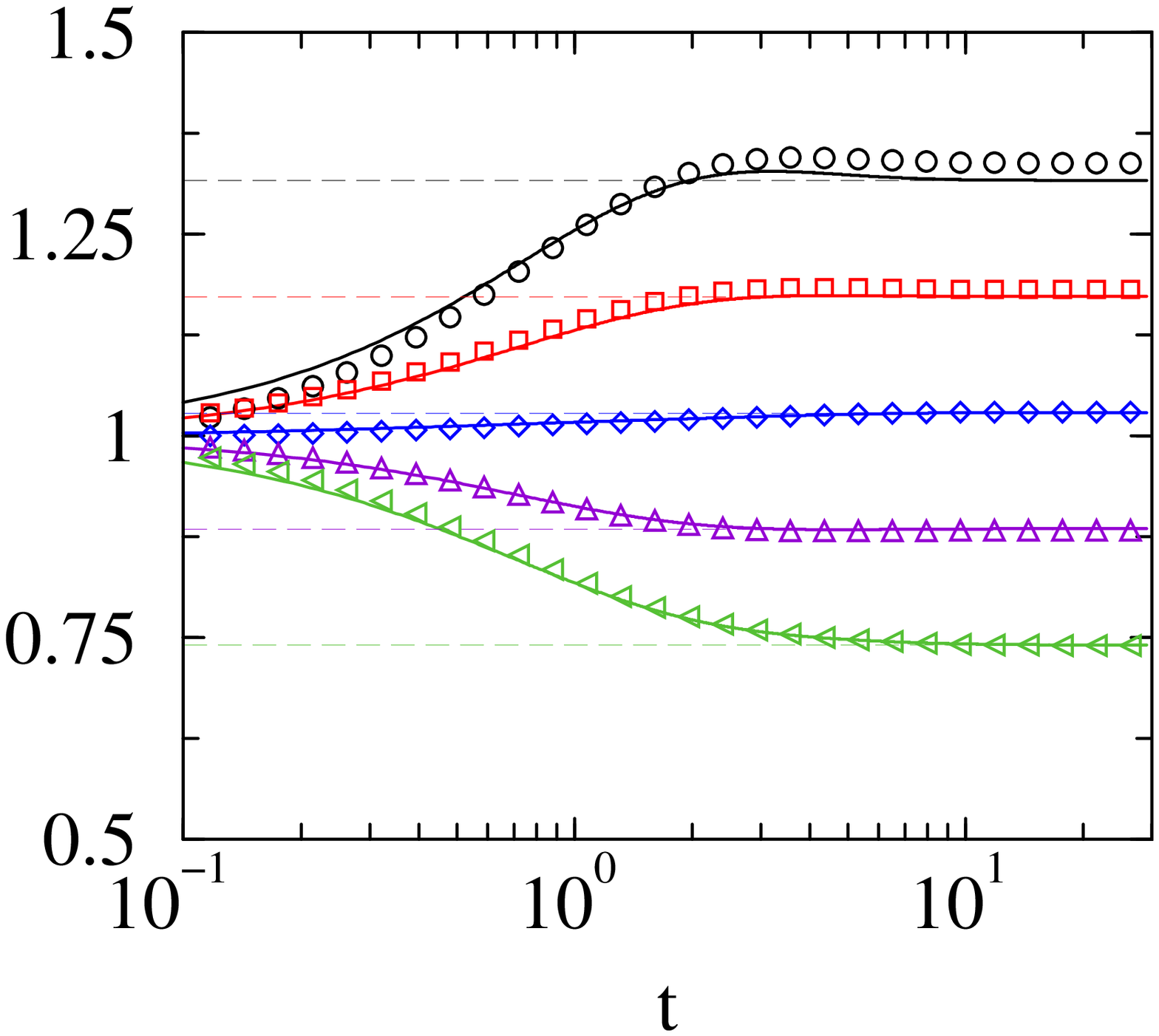}}

\put(-33,29){\large $\nu_n^{(1)}$}

\end{picture} &
\begin{picture}(100,55)
\put(-57,5){\epsfysize=48\unitlength\epsfbox{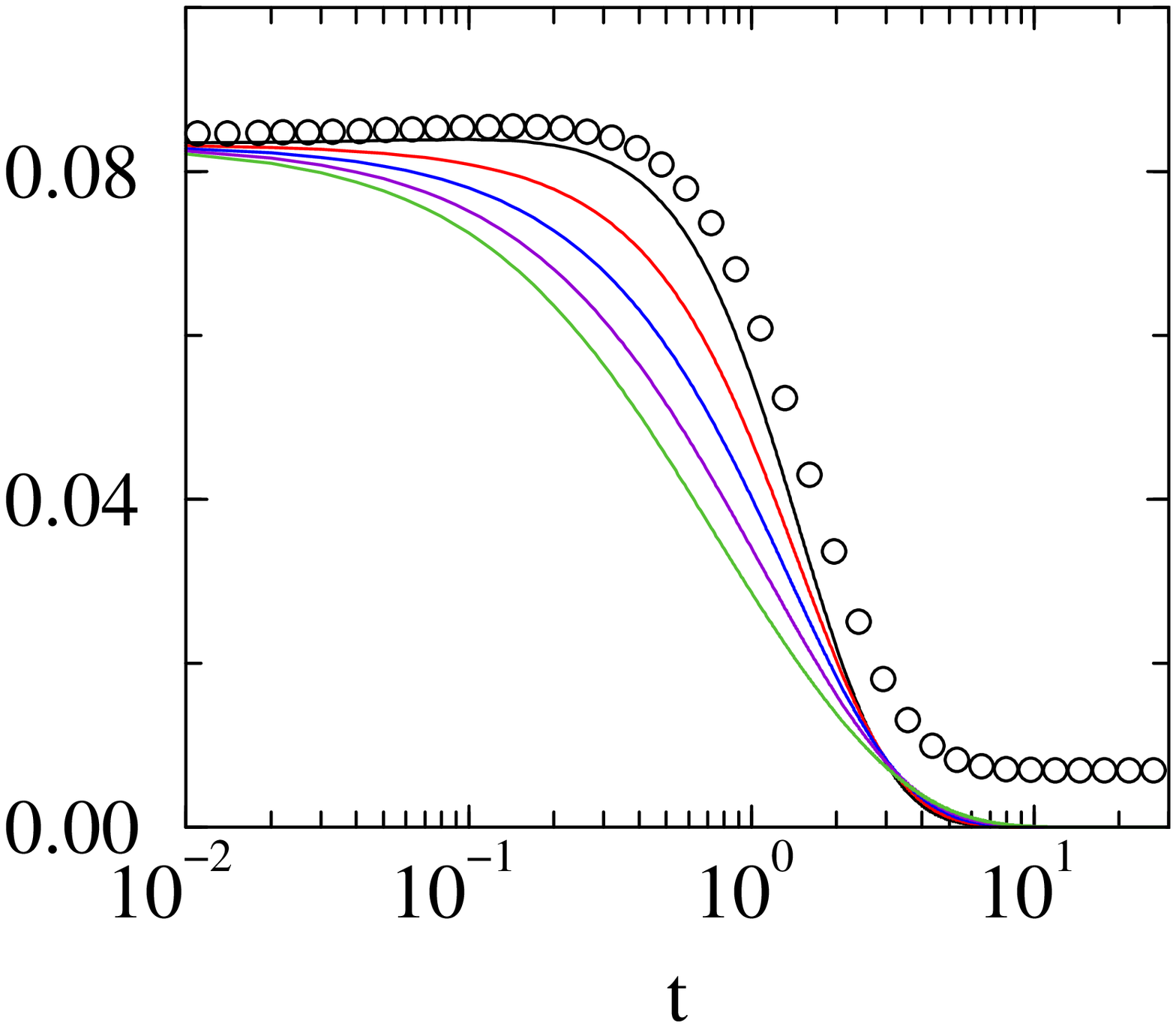}}
\put(-61,29){\large $\sigma_n^2$}
\end{picture}
\end{tabular}
\vspace*{4mm} \caption{Temporal evolution of the moments $\{\nu_n^{(1)},\sigma_n^2\}$ (see text for definitions) for the replicator system with $p=4$ traits at fixed $u=0.2,a=0.1$. Solid lines are obtained from a numerical solution of Eqs. (\ref{eq:lu},\ref{eq:la},\ref{eq:moftf}) (time-stepping $dt=10^{-6}$). Different lines are for $n=0,1,\dots,4$ from top to bottom in the left panel and from top to bottom at intermediate times in the right panel. Horizontal lines in the left panel are the fixed point values given by Eq. (\ref{fpp}). Symbols are from simulations of the replicator system with $p=4$ patterns, $N=300$, averages over $50$ samples of the disorder are taken (with circles, squares, diamonds, up and left triangles corresponding to $n=0,1,\dots,4$ respectively; numerical results for the variance (right panel) are shown only for $n=0$). Initial concentrations are drawn from a flat distribution over the interval $[0.5,1.5]$, corresponding to $\nu_n^{(1)}(t=0)=1$ and $\sigma_n^2(t=0)=0.083$ for all $n=0,\dots,4$.}
\label{fig:dynfin}
\end{figure}

In regimes in which (\ref{eq:condition}) is fulfilled the numerical
solution of (\ref{eq:lu},\ref{eq:la}) evolves towards a fixed point
$\nu^{(1)}_n=x_n>0$, $\sigma_n^2=0$ asymptotically as shown in
Fig. \ref{fig:dynfin} so that the validity of the Gaussian ansatz is
verified {\em a posteriori}. We expect the numerical solution of the
approximate equations (\ref{eq:lu},\ref{eq:la}) to capture the
dynamical behaviour of the system at least in the asymptotic regime of
large times. This is indeed confirmed by comparison with measurements
of $\nu^{(1)}_n(t)$ in numerical simulations, see the left panel of
Fig. \ref{fig:dynfin}. (We note here that the width of the
distributions $\rho_n(x)$ measured in our numerical experiments
appears to remain finite ($\sigma_n^2\approx 0.01$) up to the time
interval covered in simulations as shown in the right panel of Fig. \ref{fig:dynfin}. A similar effect was observed in
\cite{OF3}, and is presumably due to finite-size or finite
running-time effects or to discretisation errors in the numerical
scheme used to iterate the replicator equations.)

\section{Concluding remarks}
To summarise we have used dynamical techniques to study systems of
interacting replicators subject to random Hebbian couplings with and
without dilution. Our study focuses on the statistical mechanics
properties of the model and complements existing replica analyses of
the fully connected model. In the fully connected case both approaches
lead to the same equations describing the ergodic stationary states of
the system and accordingly the phase diagrams obtained from the
statics and the dynamics respectively coincide. In addition we have
identified two different types of phase transitions and patterns of
ergodicity breaking in the fully connected model, one signalled by a
diverging integrated response, the other marked by an onset of
long-term memory at finite $\chi$ as well as by a dynamical
instability of the fixed point solutions against small
perturbations. This latter dynamical transition corresponds to an
AT-instability of the replica-symmetric solution found in the statics.

The generating functional approach used here allows one to go beyond
the models accessible by static replica techniques, and we have
extended the analysis to the case of Hebbian couplings with dilution
of an arbitrary symmetry. We find that no divergencies of the
integrated response occur in diluted models, but that the transition
marked by an instability of the fixed points persists. With the scaling of
self-interactions used here dilution in general increases the range of
ergodic behaviour in the $(u,\alpha)$-plane, up to a small
region of re-entrance. The order parameters show non-trivial
behaviour as a function of the connectivity. Asymmetry in the
dilutions appears to promote an increased survival probability
of any individual species.

Finally we have presented a discussion of the dynamics of a replicator
system with a finite number of traits. Techniques developed originally
for neural networks can be used to formulate an exact hierarchical set
of equations for the concentration-densities on sub-lattices resulting
from the quenched assignment of traits. Within a Gaussian ansatz
approximate, but closed equations for the evolution of the first two
moments of these densities can be found, and allow one to study the
approach to the stationary state analytically and in good agreement
with numerical simulations.

Several extensions of the present work are possible and of potential
interest. While we have here addressed only the case of extensively
connected species the phenomenology of the the finite-connectivity
case $cN\sim{\cal O}(1)$ might be worthwhile studying with dynamical
methods now becoming available for finite-connectivity systems
\cite{Hatchetal04}. Furthermore replicator systems with
heterogeneous species-dependent degree of connectivity $c_i$ could be
taken into account leading to an ensemble of effective processes
\cite{correl}, as well as the possibility of species not carrying all
traits but only a certain individual subset. Finally an `on-line'
version of the present model could be devised in discrete time,
i.e. one in which at each step only one active trait $\mu(t)$ is
relevant for the evolution of the replicator system (as opposed to the
present `batch' model in which an effective contraction over all
traits is considered).

\section*{Acknowledgements}
This work was supported by the European Community's Human Potential
Programme under contract HPRN-CT-2002-00319, STIPCO. The author would
like to thank D Sherrington for bringing random replicator models to
his attention, and A C C Coolen for discussions on dynamical
methods. Stimulating interaction with M Bazan Peregrino is gratefully
acknowledged.


\section*{References}

\begin{thebibliography}{99}
\bibitem{Book1}Challet D, Marsili M and Zhang Y-C 2005 {\em Minority
    Games} (Oxford University Press, Oxford UK)
\bibitem{Book2}Coolen A C C 2005 {\em The Mathematical Theory of
    Minority Games} (Oxford University Press, Oxford UK)
\bibitem{Book3} Johnson NF, Jefferies P and Hui PM 2003 {\it Financial
    market complexity} (Oxford University Press, Oxford UK)
\bibitem{Book4} Hofbauer J, Sigmund K 1988 {\it Dynamical Systems and the Theory of Evolution} (Cambridge University Press, Cambridge UK)
\bibitem{Book5} Peschel M, Mende W 1986 {\it The Prey-Predator Model} (Springer Verlag, Vienna)
\bibitem{DO}
Diederich S, Opper M 1989 {\em Phys. Rev. A} {\bf 39} 4333 
\bibitem{OD}
Opper M, Diederich S  1992 {\em Phys. Rev. Lett.} \textbf{69} 1616 
\bibitem{OD2}
Opper M, Diederich S 1999 {\em Comp. Phys. Comm.} \textbf{121-122} 141 
\bibitem{OF1} 
de Oliveira V Fontanari J 2000 {\em Phys. Rev. Lett.} \textbf{85} 4984 
\bibitem{OF2} 
de Oliveira V and Fontanari J 2001 {\em Phys. Rev. E} \textbf{64} 051911 
\bibitem{OF3} 
de Oliveira V and Fontanari J 2002 {\em Phys. Rev. Lett.} \textbf{89} 148101
\bibitem{O}
de Oliveira V 2003 {\em Eur. Phys. J. B} \textbf{31} 259 
\bibitem{SF} Santos D, Fontanari J 2004  {\em Phys. Rev. E} {\bf 70} 061914
\bibitem{Tokita} Tokita K 2004 {\em Phys. Rev. Lett.} {\bf 93} 178102
\bibitem{ParisiBiscari} Biscari P, Parisi G 1995 {\em J. Phys. A: Math. Gen.} {\bf 28} 4697
\bibitem{TokitaYasumoti} Tokita K, Yasumoti A 1999 {\em Phys. Rev. E} {\bf 60} 842
\bibitem{ChaTok} Chawanya T, Tokita K 2002 {\em J. Phys. Soc. Japan} {\bf 71} 429

\bibitem{Cool00a} Coolen A C C 2001, in Handbook of Biological Physics
Vol 4 (Elsevier Science, eds F Moss and S Gielen) 513, {\tt cond-mat/0006010}

\bibitem{Cool00b} Coolen A C C 2001,  in Handbook of Biological Physics
Vol 4 (Elsevier Science, eds F Moss and S Gielen) 597, {\tt cond-mat/0006011}

\bibitem{Rieger} Rieger H 1989 {\em J. Phys. A: Math. Gen.} {\bf 22} 3447 

\bibitem{dD}
De Dominicis C, Phys. Rev. B \textbf{18}, 4913 (1978)

\bibitem{Crisanti} Crisanti A, Horner H, Sommers H-J 1193 {\em Z. Phys. B.} {\bf 92} 257

\bibitem{EO} 
Eissfeller H and Opper M 1992 {\em Phys. Rev. Lett.} {\bf 68} 2094 
\bibitem{corr} Heimel JAF, De Martino A 2001 {\em J. Phys. A: Math. Gen.} {\bf 34} L539
\bibitem{dilute} Galla T 2005 {\em J. Stat. Mech.} P01002
\bibitem{myother} Galla T 2005, preprint {\tt cond-mat/0508174}
)
\bibitem{Hatchetal04} Hatchett J P L, Wemmenhove B, Perez Castillo I,
Nikoletopoulos T, Skantzos N S, Coolen A C C 2004 {\em J. Phys. A: Math. Gen.} {\bf 37} 6201

\bibitem{correl} Galla T, Sherrington D 2005 {\em Eur. J. Phys. B} {\bf 46} 153 
\end{thebibliography}
\end{document}